\documentclass[aip,graphicx,amsmath, amssymb, reprint]{revtex4-1}

\usepackage{graphicx}% Include figure files
\usepackage{dcolumn}% Align table columns on decimal point
\usepackage{bm}% bold math
\usepackage[dvipsnames]{xcolor}
\usepackage[utf8]{inputenc}
\usepackage[T1]{fontenc}
\usepackage{mathptmx}
\usepackage{etoolbox}

\makeatletter
\def\@email#1#2{%
	\endgroup
	\patchcmd{\titleblock@produce}
	{\frontmatter@RRAPformat}
	{\frontmatter@RRAPformat{\produce@RRAP{*#1\href{mailto:#2}{#2}}}\frontmatter@RRAPformat}
	{}{}
}%
\makeatother

\graphicspath{{./Images/}}
\draft % marks overfull lines with a black rule on the right

\begin{document}
	
\preprint{AIP/123-QED}

\title[]{Impact of non-reciprocal interactions on colloidal self-assembly with tunable anisotropy}
\author{Salman Fariz Navas}
\email[]{salman.fariz.navas@tu-berlin.de}

\author{Sabine H. L. Klapp}
\email[]{sabine.klapp@tu-berlin.de}
\affiliation{Institute for Theoretical Physics, Technical University of Berlin, Hardenbergstr. 36, 10623 Berlin, Germany}

\date{\today}% It is always \today, today,

\begin{abstract}
Non-reciprocal (NR) effective interactions violating Newton’s third law occur in many biological systems, but can also be engineered in synthetic, colloidal systems. Recent research has shown that such NR interactions can have tremendous effects on the overall collective behaviour and pattern formation, but can also influence aggregation processes on the particle scale. Here we focus on the impact of non-reciprocity on the self-assembly of an (originally passive) colloidal system with \textit{anisotropic} interactions whose character is tunable by external fields. In the absence of non-reciprocity, that is, under equilibrium conditions, the colloids form square-like and hexagonal aggregates with extremely long life times yet no large-scale phase separation [Kogler et al., \textit{Soft Matter} \textbf{11}, 7356 (2015)], indicating kinetic trapping. Here we study, based on Brownian Dynamics (BD) simulations in 2D, a NR version of this model consisting of two species with reciprocal isotropic, but NR anisotropic interactions. We find that NR induces an effective propulsion of particle pairs and small aggregates (“active colloidal molecules”) forming at initial stages of self-assembly, an indication of the NR-induced non-equilibrium. The shape and stability of these initial clusters strongly depends on the degree of anisotropy. At longer times we find, for weak NR interactions, large (even system-spanning) clusters where single particles can escape and enter at the boundaries, in stark contrast to the small rigid aggregates appearing at the same time in the passive case. ln this sense, weak NR shortcuts the aggregation. Increasing the degree of NR (and thus, propulsion), we even observe large-scale phase separation if the interactions are weakly anisotropic. In contrast, system with strong NR and anisotropy remain essentially disordered. Overall, the NR interactions are shown to destabilize the rigid aggregates interrupting self-assembly and phase separation in the passive case, thereby helping the system to overcome kinetic barriers.
\end{abstract}

\maketitle

\section{\label{sec:introduction}Introduction}
In equilibrium soft-matter systems, interactions between constituents are, by definition, reciprocal, that is, they obey Newton’s third law (action-reaction symmetry) \cite{Israelachvili2011Intermolecular, praprotnik2008multiscale}. This holds even for coarse-grained descriptions where the environment is treated only implicitly. However, the action-reaction symmetry can be broken \cite{ivlev2015statistical} when the system is out of equilibrium, as it occurs in many biological systems \cite{feichtmayer2017antagonistic, long2001antagonistic, theveneau2013chase} and synthetic colloidal systems with certain “activity” or external driving \cite{singh2017non, grauer2021active, valadares2010catalytic, niu2018dynamics, schmidt2019light, meredith2020predator}. Effective descriptions of such systems often involve non-reciprocal (NR) interactions. An example is the following: Consider a binary mixture of colloids, where one species (say, species B) can chemically react with the surrounding solvent, resulting in the formation of certain products, while the other (say, species A) remain inert with respect to the solvent. This can be achieved, for e.g., by coating the B-particles with a catalyst. The A-particles, although not directly coupled to the solvent, can respond to the chemical gradient caused by the reaction products
(due to diffusiophoresis \cite{anderson1986transport, anderson1989colloid}) and thus move towards regions with high product concentrations. Ultimately, when particles of opposite species approach each other, the (inert) A-particle is attracted to the (reactive) B-particle, but not vice-versa \cite{reigh2018diffusiophoretically, schmidt2019light, sturmer2019chemotaxis, ruckner2007chemically, reigh2015catalytic}. 

Recent research has shown that NR interactions can have crucial impact on the collective behaviour and pattern formation \cite{fruchart2021non} in systems of active particles and scalar mixtures \cite{fruchart2021non, dinelli2023non, you2020nonreciprocity, kreienkamp2022clustering}. In the present paper we are specifically concerned with self-assembly processes on the particle scale. In particular, we ask for the impact of NR interactions on cluster aggregation of passive (i.e., non-motile) colloidal particles with \textit{anisotropic} interactions.

A variety of studies involving systems with \textit{isotropic} NR interactions have already shown interesting effects such as the formation of "active colloidal molecules" with self-propulsion in theory \cite{fehlinger2023collective, soto2014self, soto2015self, ai2023brownian, varma2018clustering, bartnick2016emerging} and in experiments \cite{niu2018dynamics, schmidt2019light, grauer2021active, ruckner2007chemically, valadares2010catalytic, reigh2015catalytic}. In some of these systems, NR even induces phase separation \cite{agudo2019active, chiu2023phase}. More generally, NR has found to have profound effects on complex (e.g., multifarious) self-assembly where, in equilibrium, the aggregation process is often stuck in metastable minima of the free energy landscape, that is, the system is kinetically trapped. In such situations, NR can help to overcome kinetic barriers \cite{osat2023escaping, osat2023non, klapp2023non, bartnick2016emerging}, similar to what is seen in complex self-assembly processes of Janus colloids when these are activated, e.g., by light \cite{mallory2019activity}.

In this study, we are specifically interested in the interplay of NR, on the one hand, and anisotropy of interactions, on the other hand. To this end we investigate a NR version of a two-dimensional (2D) model of a (passive) colloidal model system \cite{kogler2015generic} describing aggregation in orthogonal electric and magnetic fields \cite{bharti2015assembly}. The fields are assumed to induce four “patches” on the particles whose detailed configuration depends on the field strength \cite{kogler2015generic}. The resulting anisotropy of the two-particle interaction thereby becomes tunable, allowing for different cluster types from hexagonal (low anisotropy) to square-like (strong anisotropy). Such clusters dominate the equilibrium collective behaviour over a broad range of temperatures and densities; in particular, there is no large-scale phase separation despite the fact that the anisotropic interactions are, overall, attractive \cite{kogler2015generic}. Here we extend this model by introducing two species of particles whose inter-species anisotropic interactions are NR (the isotropic part is symmetric). We show that already small degrees of NR lead to effective propulsion on the two-particle level and to activity of small aggregates formed at initial stages of aggregation. This has profound impact on the speed of aggregation and also on the system’s long-time behaviour. In particular, we observe an analogue of motility-induced phase separation familiar from active particles at large NR and low anisotropy. In contrast, strong anisotropy destroys this effect.

The rest of the paper is organised as follows. In Section~\ref{sec:model}, we summarize features of the original passive model \cite{kogler2015generic} and introduce the NR variant. Our numerical results are described in Sec.~\ref{section: results}, focussing first on the dynamics of a NR pair of particles. We then move to aggregation in weakly NR systems as compared to that in the reciprocal reference case, and eventually to strongly NR systems. Finally, in Section~\ref{section: conclusion}, we summarize our findings.

\section{\label{sec:model}Model}

\begin{figure*} [htp]
	\centering 
	\includegraphics[width=1.0\textwidth]{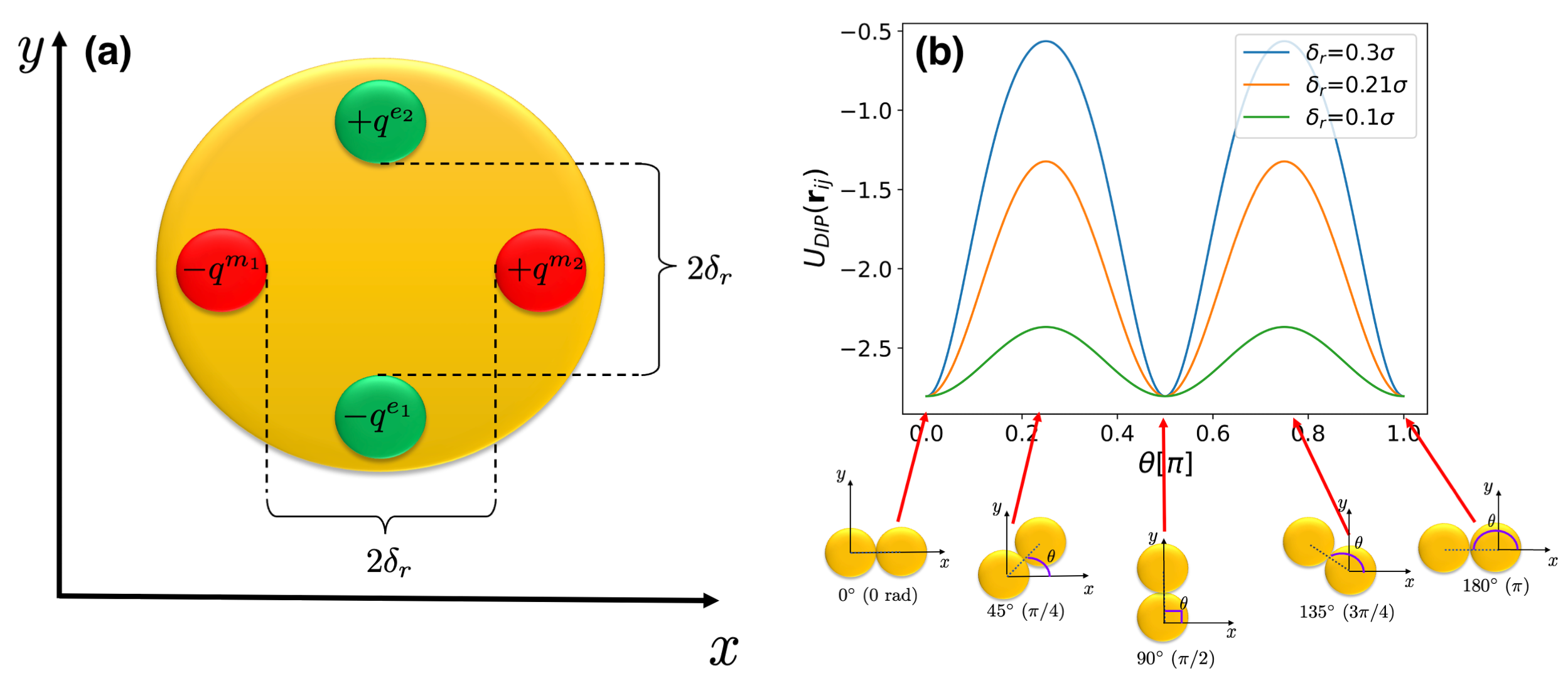}
	\caption{(a) Schematics of a particle with two pairs of fictitious point charges (mimicking induced dipole moments) separated by a distance of $2\delta_r$. The point charges are colour coded according to the corresponding dipole type: green for electric field ($e$) pointing in $y-$direction and red for the magnetic field ($m$) pointing in $x-$direction. (b) Anisotropic potential $U_{\text{DIP}}(\mathbf{r}_{ij})$ for two particles at contact ($r_{ij}=\sigma$) and at different spatial configurations plotted as a function of the angle $\theta$ between the connecting vector $\mathbf{r}_{ij}$ and the $x-$axis. In the bottom, we show typical configurations pertaining to specific values of $\theta$ (indicated by red arrows). The black arrows indicate the $x-$ and $y-$axes. One particle is kept in place, while the other is rotated around that particle. The dashed line represents the connecting vector.} \label{Fig:1}
\end{figure*}

Our model is inspired by previous work on two-dimensional (2D) systems of colloidal particles in orthogonal electric ($e$) and magnetic ($m$) fields \cite{kogler2015generic}. Specifically, we consider a system of $N$ disk-like particles with diameter $\sigma$, where we distinguish two species, A and B with $N_A=N_B=N/2$. The intra-species (A-A and B-B) interactions are reciprocal and equal to each other regarding both, isotropic (steric) and orientational, contributions. The same holds for the isotropic part of the inter-species (A-B and B-A) interactions. In contrast, the orientational inter-species interactions are NR. 

To introduce the details of the orientational pair interactions, we start by describing the effect of the external fields on one particle (of either species), for details see Ref. \cite{kogler2015generic}. We assume that the external fields of type $m$ and $e$ are pointing along the $x-$ and $y-$axes, respectively, and induce two dipole moments along their respective directions in each particle \cite{bharti2015assembly, bharti2016multidirectional, kogler2015generic}. The two dipole moments are assumed to be of equal magnitudes and to be independent of each other, i.e., they interact only with dipole moments of the same type in other particles. We mimic each of these dipole moments using two fictitious, opposite point charges $q^{\alpha_1}=-q^{\alpha_2}$ separated by a distance $2\delta_r$, where index $\alpha=e, m$, represents the two fields. A schematic of the particle is given in Fig.~\ref{Fig:1}(a). The displacement vector of the point charges from the particle's centre is given by $\boldsymbol{\delta^{\alpha_k}}=(-1)^k\delta_r\mathbf{\hat{e}_\alpha}$, where the index $k=1,2$ and the unit vectors $\mathbf{\hat{e}_e}=\mathbf{\hat{y}}$ and $\mathbf{\hat{e}_m}=\mathbf{\hat{x}}$. The parameter $\delta_r$ describes the distance between each of these point charges from the particle’s centre and may be considered as a measure of the strength of the fields. Allowing the two dipole moments to have different magnitudes will expand the parameter space leading to separate ${\delta_r}_x$ and ${\delta_r}_y$  along the $x-$ and the $y-$ axes. The resulting interactions will then depend on the ratio ${\delta_r}_x/{\delta_r}_y$. We note that the limiting cases (when ${\delta_r}_x\gg {\delta_r}_y$ or ${\delta_r}_y \gg {\delta_r}_x$) of having only one dipole moment are well studied (for e.g., see \cite{kretschmer2004pearl, wang2013magnetic, erb2009magnetic}). We thus feel that setting both as equal provides a particularly interesting case. Following \cite{kogler2015generic}, we assume that each point charge has the same absolute value $|\pm q|=2.5\left(\epsilon/\sigma\right)^{1/2}$. The interaction potential between two particles of the same species $\beta$ (with $\beta=A, B$) is given by 

\begin{equation}
	U_{\beta \beta}(\mathbf{r}_{ij})=U_{\text{WCA}}(r_{ij})+U_{\text{DIP}} (\mathbf{r}_{ij}) \label{eq:U}.
\end{equation}

The first term on the right side of Eq.~(\ref{eq:U}) represents the steric repulsion of the particles and is described by the Weeks-Chandler-Andersen (WCA) potential  \cite{wca}, $U_{\text{WCA}}(r_{ij})=4\epsilon\left[    \left( \frac{\sigma}{r_{ij}} \right)^{12} - \left( \frac{\sigma}{r_{ij}} \right)^6+ \frac{1}{4} \right]$, where $\epsilon$ is the WCA energy parameter that is used as the unit for energy for our simulations, and $r_{ij}=|\mathbf{r}_{ij}|=|\mathbf{r}_i-\mathbf{r}_j|$ is the distance between the particle pairs. The potential $U_{\text{WCA}}(r_{ij})$ is made purely repulsive by employing a cut-off distance of $r_{\text{WCA}}=2^{1/6}\sigma$. The second term on the right side of Eq.~(\ref{eq:U}) describes the interaction potential due to the two point charge pairs on each particle. We implicitly assume that the interactions between these point charges are screened by choosing a Yukawa potential\cite{kogler2015generic}. The interaction between two point charges of different particles $i, j$ of type $\alpha=\text{e, m}$ and sign $k,l=1, 2$ is described by, $U^{\alpha_{k l}}_{\text{YU}}({r_{ij}^{\alpha_{kl}})}=q^{\alpha_k} q^{\alpha_l} \frac{\exp\left(-\kappa r_{ij}^{\alpha_{kl}}\right)}{r_{ij}^{\alpha_{k l}}}$, where $r_{ij}^{\alpha_{k l}}=|\mathbf{r}_{ij}+\boldsymbol{\delta}^{\alpha_l}-\boldsymbol{\delta}^{\alpha_k}|$  is the distance between the charge pairs. The overall contribution of the charges to the pair-interaction potential is given by the sum of the Yukawa potentials for the magnetic and electric charges, $	U_{\text{DIP}}(\mathbf{r}_{ij})=\sum_{\alpha\in \text{e,m}}\sum_{k,l=1}^{2} U^{\alpha_{kl}}_{\text{YU}}({r_{ij}^{\alpha_{kl}})}$. In the present study we consider three different values of $\delta_r$, $0.1\sigma$, $0.21\sigma$ and $0.3\sigma$. The reasoning behind the choice of these three values of $\delta_r$ is discussed later in Section \ref{Section: aggregation_behavior} and Appendix \ref{appendix: delr_threshold}. The inverse screening length $\kappa$ is  set to $\kappa=4.0/\sigma$. The potential is cut-off at a distance $r_{\text{YU}}=4.0\sigma$.   

In Fig.~\ref{Fig:1}(b), we plot the potential $U_{\text{DIP}}(\mathbf{r}_{ij})$ for two particles that are in contact with each other as function of $\theta$, the polar angle between $\mathbf{r}_{ij}$ and the $x-$axis, for different values of charge separation $\delta_r$. It is seen that $U_{\text{DIP}}(\mathbf{r}_{ij})$ is negative for all values of $\theta$ at the distance considered and also for all $\delta_r$. This shows that $U_{\text{DIP}}(\mathbf{r}_{ij})$ is, on the angular average, purely attractive. Furthermore, $U_{\text{DIP}}(\mathbf{r}_{ij})$ has minima at $\mathbf{r}_{ij}=\sigma \hat{\mathbf{e}}_{\alpha}$ (with $\alpha=\text{e, m}$), i.e., when the particles are aligned along the direction of one of the external fields, reflecting the direction dependency of the potential. Increasing $\delta_r$ renders the minima of $U_{\text{DIP}}(\mathbf{r}_{ij})$ more pronounced, resulting in an increase of anisotropy. In conclusion, when two particles get in contact with each other, they feel an (angle-dependent) mutual attraction. As a consequence, at sufficiently large coupling strength, the particles tend to stick to each other, forming a simple example of a "colloidal molecule" \cite{shen2016designing, lowen2018active}. Note that we have normalised  $U_{\text{DIP}}(\mathbf{r}_{ij})$ such that it has a constant value for $r_{ij}=\sigma$ and $\mathbf{r}_{ij}$ pointing along one of the fields ($\mathbf{r}_{ij}=\sigma\hat{\mathbf{e}}_{\alpha}$) (see Appendix \ref{appendix: U_dip_norm}), independent of $\delta_r$.

Now, we turn to the two-species system. The NR is introduced for the inter-species orientational interactions (A-B and B-A). Specifically, we modify the interaction potential in Eq.~(\ref{eq:U}) by introducing a pre-factor $D_{\beta \gamma}$ in front of $U_{\text{DIP}}(\mathbf{r}_{ij})$, where the indices  $\beta, \gamma \in \left\{ A,B\right\}$  correspond to the particle species and $D_{\beta\gamma}$ are the elements of a $2\times2$ NR matrix defined by 

\begin{equation}
	\mathbf{D}=
	\begin{pmatrix}
		1 & 1+\kappa_{\text{nr}} \\
		1-\kappa_{\text{nr}} & 1 
	\end{pmatrix}. \label{eq:non_rp_matrix}
\end{equation}

As seen from Eq.~(\ref{eq:non_rp_matrix}), the diagonal elements $D_{AA}=D_{BB}=1$, while the off-diagonal elements relate to the inter-species interactions and can differ depending on the NR parameter $\kappa_{\text{nr}}$ ($\kappa_{\text{nr}}\neq0$ implies a NR system). We here consider system with $\kappa_{\text{nr}}\in[0, 1]$, such that the sign of the B-A interaction potential remains unchanged. Note that the first term in Eq.~(\ref{eq:U}), $U_{\text{WCA}}(r_{ij})$, remains reciprocal. The complete interaction potential for two particles $i$ and $j$ belonging to species $\beta$ and $\gamma$ respectively is then given by, 

\begin{equation}
	U_{\beta \gamma}(\mathbf{r}_{ij})=U_{\text{WCA}}(r_{ij})+ D_{\beta \gamma}U_{\text{DIP}} (\mathbf{r}_{ij}) \label{eq:U_nr}.
\end{equation} 

The motion of the particles is governed by the overdamped Langevin equation, 

\begin{equation}
	\gamma\dot{\mathbf{r}}_{i,\beta}=-\sum_{j\neq i}\nabla_{\mathbf{r}_{i}} U_{\beta\gamma}(\mathbf{r}_{ij})+\mathbf{F}_{R,i}(t), \label{eq:BD}
\end{equation} 

where $\mathbf{r}_{i, \beta}$ is the position vector of each particle $i$ of species $\beta=A, B$, with $i=1,..,N$ and  $\mathbf{r}_{ij}=|\mathbf{r}_{i, \beta}-\mathbf{r}_{j,\gamma}|$ is the centre-to-centre displacement vector between particle $i$ and another particle $j$ of species $\gamma=A,B$. $\mathbf{F}_{R,i}(t)$ is a random force acting on particle $i$ at time $t$ with the properties of a Gaussian white noise, that is, $\left< F^a_{R,i}(t) \right>=0$ and $\left<F^a_{R,i}(t)  F^b_{R,j}(t') \right>=2\gamma k_BT \delta(t-t')\delta_{ij}\delta_{ab}$ where $i, j$ are particle indices and $a,b \in \left\{ 1,2\right\}$ represent the vector components. The properties of the random force remain the same for both species of particles. Further, $\gamma$ is the friction coefficient of the medium in which the colloidal particles are dispersed in, $T$ is the temperature and $k_B$ the Boltzmann constant. 

We perform simulations consisting of an equal number of particles of both species, $N_A=900$ and $N_B=900$, for a total of $N=N_A+N_B=1800$ particles in a square-like simulation box with periodic boundary conditions. The reduced number density $\rho^*=\rho\sigma^2$ is set to 0.3 and the reduced temperature is set to $T^*=k_BT/\epsilon=0.05$. The temperatures and densities are chosen such that the system remains below the crystallization and the anisotropic interactions remain dominant. Simulations are performed for different values of $\kappa_{\text{nr}}$, ranging from 0 to 1. The Langevin equations~(\ref{eq:BD}) are solved via the Euler-Maruyama integration scheme for stochastic ordinary differential equations\cite{kloeden1992stochastic}. The integration step-size is set to $\Delta t=10^{-5}\tau_B$, where $\tau_B$ is defined as the Brownian timescale of the system and is given as $\tau_B= \sigma^2\gamma/k_BT$. The simulations are performed until the simulation time reaches $100\tau_B$. 

\section{Results} \label{section: results}
\subsection{Dynamics of two particles}  \label{section: nrp_propulsion}
As a first step to understand the collective behaviour resulting from Eqs.~(\ref{eq:U_nr}) and (\ref{eq:BD}), we consider the dynamics of two nearby particles of different species. The anisotropic characteristics of $U_{\text{DIP}}(\mathbf{r}_{ij})$ alone, i.e., without the NR pre-factor $D_{\beta \gamma}$ in Eq.~(\ref{eq:U_nr}), are summarized in Sec.~\ref{sec:model}. We now explore the effect of the full NR potential given in Eq.~(\ref{eq:U_nr}). 

\begin{figure*} [htp]
	\centering 
	\includegraphics[width=1.0\textwidth]{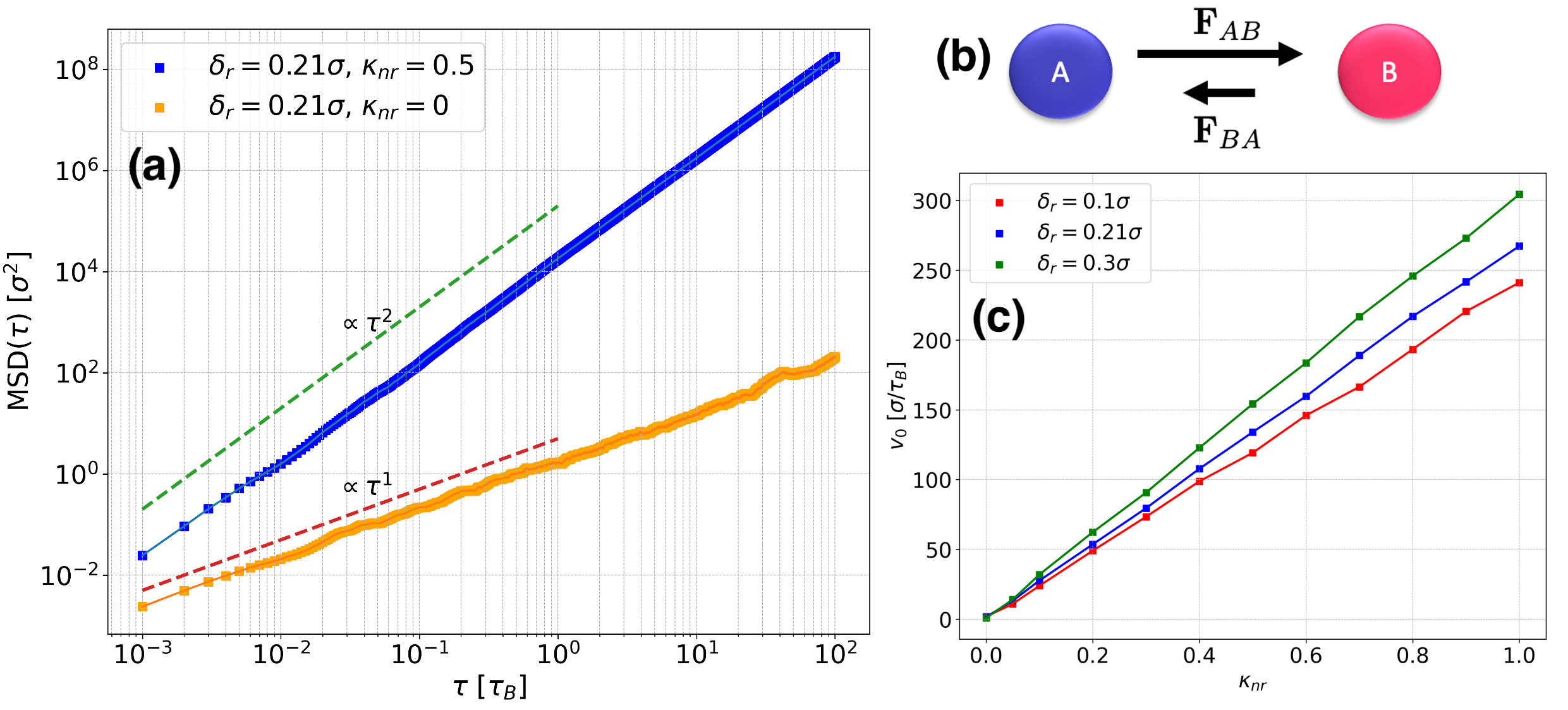}
	\caption{(a) MSD of a particle pair at contact with reciprocal (orange) and NR (blue) interactions. The green and red dashed lines indicate ballistic ($\text{MSD}\propto\tau^2$) and diffusive ($\text{MSD}\propto\tau$) behaviour, respectively. (b) Schematics of a non-reciprocally interacting particle pair indicating the mutual forces. (c) Plot of the effective propulsion velocity $v_0$ of the NR particle pair as a function of $\kappa_{\text{nr}}$ for the different values of $\delta_r$. The simulations are performed at $T^*=0.05$.} \label{Fig:2}
\end{figure*}

If $\kappa_{\text{nr}}>0$, it follows from Eq.~(\ref{eq:U_nr}) that $|U_{AB}(\mathbf{r}_{ij})|>|U_{BA}(\mathbf{r}_{ij})|$, for all distances and angles between the A- and B- particle. This means that the force $\mathbf{F}_{AB}=-\nabla_{\mathbf{r}_{i}}U_{AB}$ acting on an $A-$ particle due to a neighbouring B- particle is more attractive than the other way around. This situation is depicted schematically in Fig.~\ref{Fig:2}(b). Note that the steric repulsions are reciprocal in nature for all combinations of particles (this choice is similar to that in other studies involving NR Brownian systems \cite{mandal2022robustness, fehlinger2023collective}). Therefore, the overall effect when a particle of species A gets close to one of species B is that they tend to stick together because of the attractive nature of the anisotropic potential, but A tends to push B along the direction of attraction. This remarkable effect can be seen as a propulsion mechanism, resulting in a directed motion of the pair of particles considered. 

In order to quantify the degree of propulsion, we perform simulations of a dilute system consisting of one A-B pair of particles (that are initially at contact) for $100\tau_B$  and calculate the mean squared displacement (MSD) defined as

\begin{equation}
	\text{MSD}(\tau)=\sum_{i=1}^{2}\left<\left[\mathbf{r}_i(t+\tau)-\mathbf{r}_i(t)\right]^2 \right>/2. \label{eq: MSD_2}
\end{equation}

Here, the $\left<..\right>$ brackets denote a noise as well as time average and the factor of $1/2$ is introduced to average over the two particles. We average over 50 noise realisations for each parameter combination to calculate the MSD. Two exemplary MSDs illustrating the reciprocal ($\kappa_{\text{nr}}=0$) and NR ($\kappa_{\text{nr}}>0$) case are plotted in Fig.~\ref{Fig:2}(a). In both cases the two particles stick together for the entire run. However, the time dependence is different. In particular, the MSD of the non-reciprocally interacting A-B pair shows a ballistic behaviour, that is, $\text{MSD}(\tau)\sim\tau^2$, for the entire length of the simulation. This reflects persistent motion. We note that the absence of a long-time diffusive regime is expected due to the absence of other particles. In contrast, the MSD of the reciprocally interacting pair shows diffusive behaviour, that is, $\text{MSD}(\tau)\sim\tau^1$. 

The MSD of the NR pair resembles that of an active Brownian particle (ABP) with an infinite persistence time \cite{howse2007self, ten2011brownian,volpe2014simulation}. This motivates us to fit this MSD to the analytic expression for the MSD of an ABP \cite{howse2007self, ten2011brownian, zottl2016emergent} in the ballistic regime (see Appendix \ref{appendix: v_0_fit}). From the fit parameters, we can extract the effective propulsion velocity $v_0$. In Fig.~\ref{Fig:2}(c), we present $v_0$ as a function of the NR parameter $\kappa_{\text{nr}}$ for different values of $\delta_r$. It can be seen that self-propulsion is induced as soon as $\kappa_{\text{nr}}>0$ [indicated by the non-zero value of $v_0$ at $\kappa_{\text{nr}}=0.05$ in Fig.~\ref{Fig:2}(c)]. Further, $v_0$ increases with $\kappa_{\text{nr}}$ for all $\delta_r$ values considered, clearly revealing the role of NR for the existence and strength of propulsion. Another interesting observation is that $v_0$ also depends on $\delta_r$. At fixed $\kappa_{\text{nr}}$, $v_0$ increases with $\delta_r$, and this effect becomes more pronounced the larger $\kappa_{\text{nr}}$ is. This can be attributed to the direction-dependent nature of the potential, whose minima become more pronounced with increasing $\delta_r$. 

To summarize, the non-reciprocally interacting pair can be seen as a realization of a simple active colloidal "molecule" \cite{lowen2018active, grauer2021active} (specifically, a dimer). However, we emphasize that the "activity" in our system is solely due to NR inter-species interactions. This is different from studies where the individual particles are intrinsically  self-propelled \cite{vicsek1995novel, toner1998flocks,sturmer2019chemotaxis, golestanian2012collective, fily2012athermal, singh2017non, sharifi2016pair, hauke2020clustering, vuijk2022active, knevzevic2022collective}. Similar propulsion mechanisms induced by NR interactions have also been reported in other recent studies, both in theory \cite{fehlinger2023collective, soto2014self, soto2015self, ai2023brownian, varma2018clustering, bartnick2016emerging} and in experiments \cite{schmidt2019light, grauer2021active, ruckner2007chemically, valadares2010catalytic, reigh2015catalytic}, yet for isotropic interactions. 

\subsection{Aggregation in reciprocal and weakly non-reciprocal systems}  \label{Section: aggregation_behavior}

\begin{figure*} [htp]
	\centering
	\includegraphics[width=1.0\textwidth]{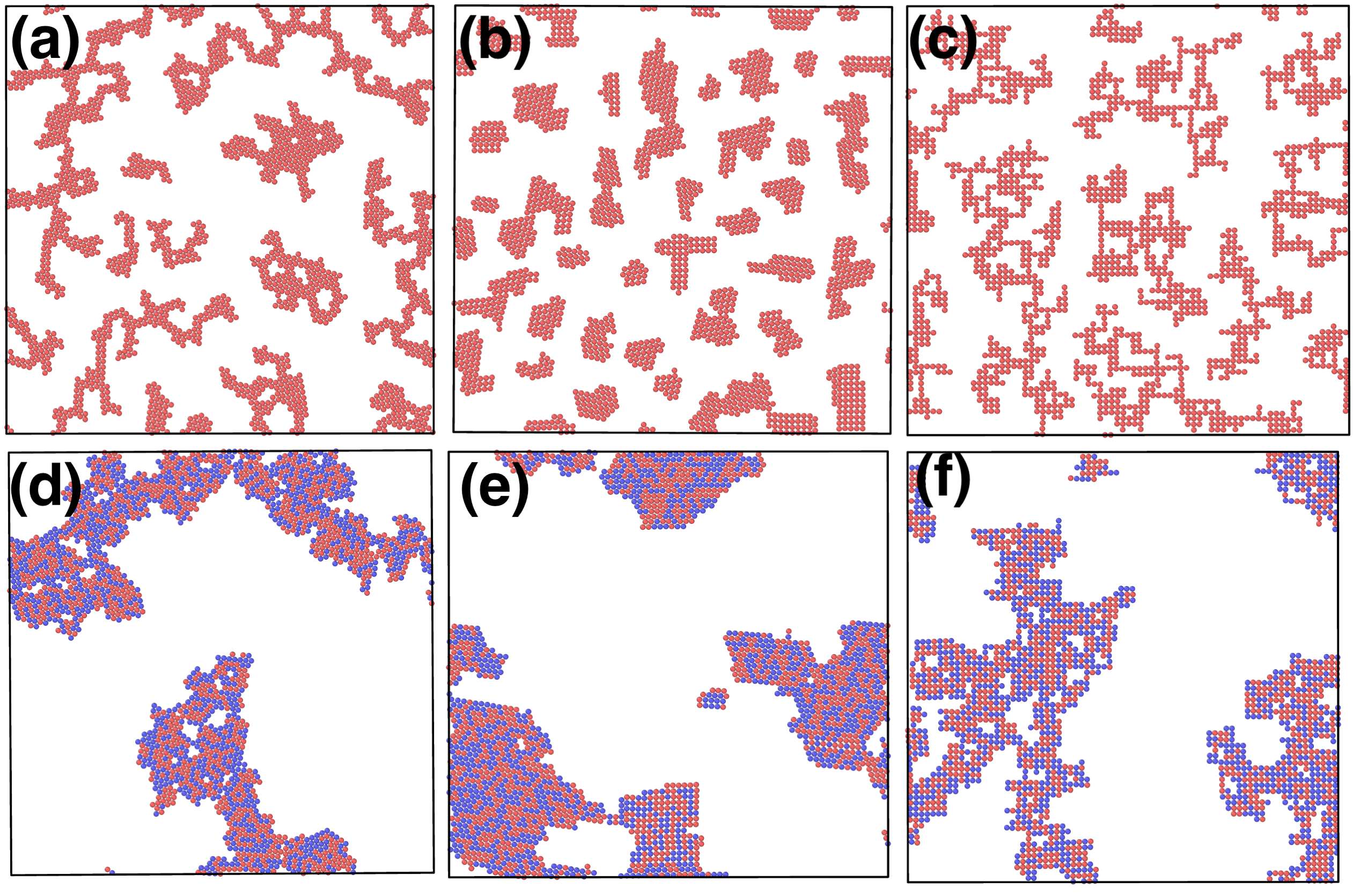}
	\caption{Snapshots of simulations started from random, uniformly distributed particles at $T^*=0.05$ and $\rho^*=0.3$ after $t=100\tau_B$ for the single-species (reciprocal) system with (a) $\delta_r=0.1\sigma$, (b) $\delta_r=0.21\sigma$, (c) $\delta_r=0.3\sigma$ and for the two-species (NR) system with $\kappa_{\text{nr}}=0.1$ and (d) $\delta_r=0.1\sigma$, (e) $\delta_r=0.21\sigma$ and (f) $\delta_r=0.3\sigma$.  For the two-species system, the blue particles correspond to particles of species A and the red ones to species B.} \label{Fig:3}
\end{figure*}

We now turn to the collective behaviour of the many-particle system consisting of $N=1800$, particles. In the next paragraph we first consider, as a reference, the reciprocal case ($\kappa_{\text{nr}}=0$) corresponding to a single-species system (see also Ref. \cite{kogler2015generic}). In Sections~\ref{subsection: non-reciprocal system} and~\ref{subsection: phase separation and disorder} we then explore the impact of NR ($\kappa_{\text{nr}}>0$). To quantify the self-aggregation process we study the evolution of the size of the largest cluster, $N_{max}$, as a function of time $t$. We calculate $N_{max}(t)$ based on a distance criterion. Particles whose centre-to-centre distances are smaller than $r_{cl}=1.15\sigma$ are defined as being bonded, and all particles that are mutually bonded are defined as belonging to the same cluster. We then determine the cluster with the largest number of constituent particles. The number of particles in the largest cluster yields $N_{max}$. To characterize the stability of the aggregates, we calculate the time-dependent bond auto-correlation function $B(\tau)$ (see Appendix \ref{appendix: BAC}). Further target quantities are the distribution of the position resolved local area fractions (see Appendix \ref{appendix: position resolved local area fracs}) and the MSD. 

\subsubsection{The reciprocal reference system} \label{subsection: reciprocal system}
First, we focus on the reciprocal case. It has been shown in Ref. \cite{kogler2015generic} that, at the temperature and density considered, $\delta_r$ is the key parameter that can tune the aggregation. In Figs.~\ref{Fig:3}(a)-(c) we show snapshots of simulations started from a random, uniform distribution of particles, after a simulation time of $t=100\tau_B$ for the three values of $\delta_r$ (already considered in Ref \cite{kogler2015generic}). We observe multiple aggregates in all three cases. However, the nature of these aggregates depend on $\delta_r$. For weakly anisotropic interactions [corresponding to low $\delta_r$ in the range of 0.05-0.21 $\sigma$, see the example $\delta_r=0.1\sigma$ shown in Fig.~\ref{Fig:3}(a)], hexagonal structures are preferred due to their packing efficiency. In contrast, when the interactions become strongly anisotropic [corresponding to high $\delta_r$ in the range of 0.21-0.35 $\sigma$, see the example $\delta_r=0.3\sigma$ shown in Fig.~\ref{Fig:3}(c)], quadratic and chain-like structures, aligning along the direction of the induced dipoles, are preferred. At the threshold value [see the case $\delta_r=0.21\sigma$ shown in Fig.~\ref{Fig:3}(b), as well as Appendix \ref{appendix: delr_threshold} for further details], neither the hexagonal structure, nor the quadratic structure is strongly preferred, rather there is a competition. This leads to a system where hexagonal and quadratically ordered aggregates exist simultaneously. The three values of $\delta_r$ serve as examples of systems at high and low anisotropies as well as at the transitioning case.

\begin{figure*} [htp]
	\centering
	\includegraphics[width=1.0\textwidth]{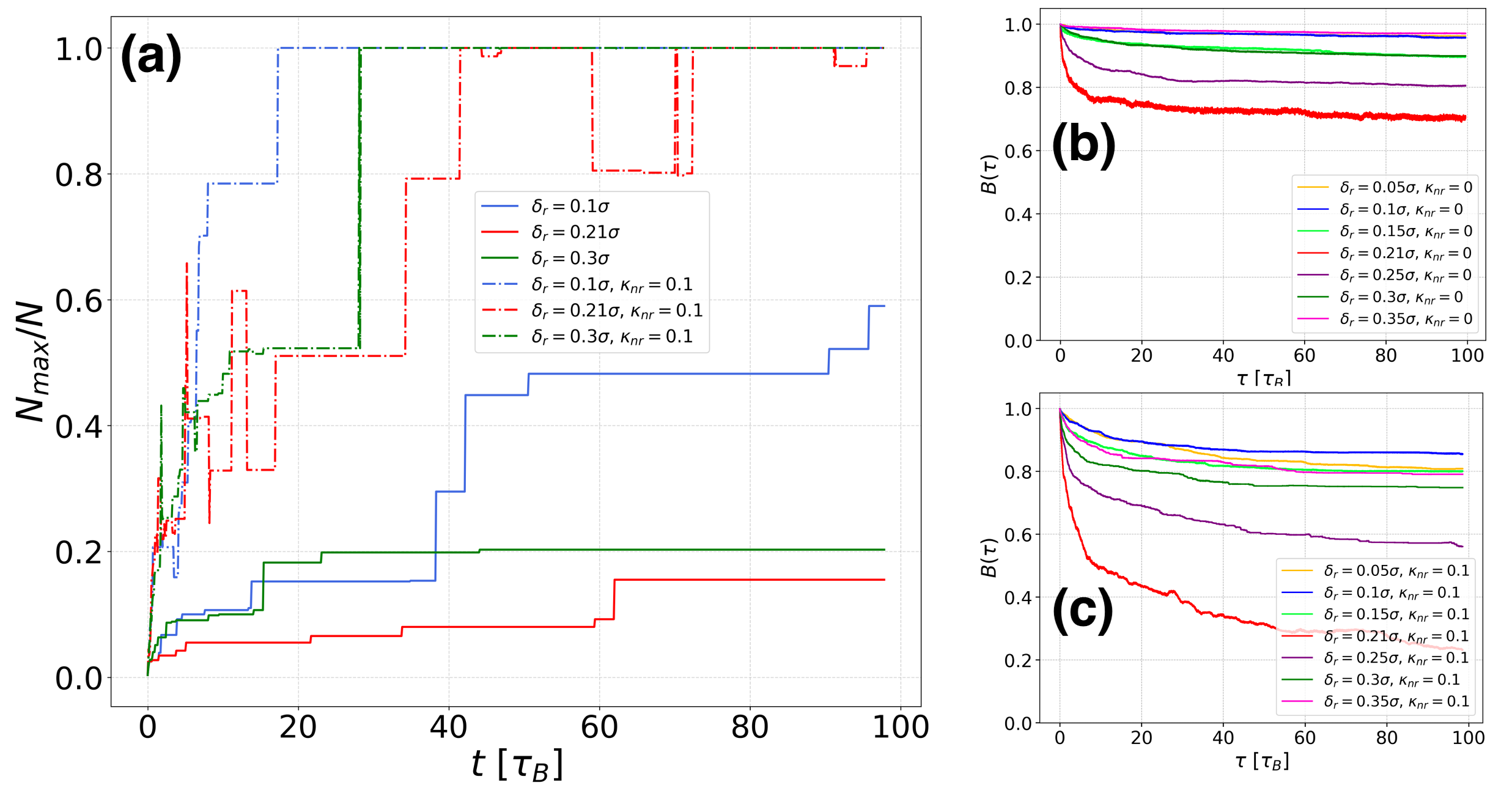}
	\caption{(a) Largest cluster size ($N_{max}/N$) as a function of time for $\delta_r=0.1\sigma, 0.21\sigma, 0.3\sigma$ in the reciprocal (solid lines) and the NR case (dashed lines) with $\kappa_{\text{nr}}=0.1$. (b) Bond auto-correlation function as a function of time difference $\tau$ for different values of $\delta_r$ in the reciprocal case and (c) in the NR case. The data pertain to $T^*=0.05$ and $\rho^*=0.3$.} \label{Fig:4}
\end{figure*}

The different types of local structure also affect the stability of the clusters. To quantify this, we plot in Fig.~\ref{Fig:4}(b) the bond auto-correlation function as a function of time difference $\tau$ for different values of $\delta_r$. Here, $B(\tau)$ is calculated starting from the finite time at which initial aggregates form ($t_0=0.5\tau_B$) until $\tau=100\tau_B$. For a system consisting of aggregates with absolutely stable, rigid bonds, we would expect the bond autocorrelation function to remain constant in time. Here, we see that $B(\tau)$ decays for all the values of $\delta_r$ in Fig.~\ref{Fig:4}(b), yet very slowly. Specifically, the decay of $B(\tau)$ is slowest for $\delta_r=0.35\sigma$ and $\delta_r=0.05\sigma$, and fastest for $\delta_r=0.21\sigma$. This indicates that the aggregates in the $\delta_r=0.05\sigma$ and $\delta_r=0.35\sigma$ cases are the most stable. The stability of the aggregates in the $\delta_r=0.05\sigma$ ($\delta_r=0.35\sigma$) case can be attributed to the fact that these systems have the weakest (strongest) anisotropic interactions among all the values of $\delta_r$ shown here. This leads to a clear preference in the orientation of the aggregates (either hexagonal or quadratic). The aggregates are the least stable at the value of $\delta_r=0.21\sigma$. The value of $\delta_r=0.21\sigma$ is the threshold value at which the aggregates in the (reciprocal or nearly reciprocal) system change from being predominantly hexagonally ordered to also show some quadratic ordering. This decrease in aggregate stability can be attributed to the competition of structures in the case $\delta_r=0.21\sigma$. For all these systems, $B(\tau)$ never reaches zero within the simulation time. 

In addition, we show in Fig.~\ref{Fig:4}(a) the quantity $N_{max}/N$ as a function of time for a particular noise realisation for the three values of $\delta_r$ [corresponding to solid lines in Fig.~\ref{Fig:4}(a)]. We stress that the general trends were found to be robust for different noise realisations. It can be seen that $N_{max}/N$ increases in a step-like manner for all three values of $\delta_r$. Remarkably, in all cases it remains below 0.6 even after $100\tau_B$. This shows that, within the simulation time, the aggregation is not complete. Additionally, it can be seen that $N_{max}$ evolves more slowly for $\delta_r=0.21\sigma$ than in the other two cases, particularly compared to the case $\delta_r=0.1\sigma$. From a physical point of view, we can understand the aggregation process in the three cases as follows. For all values of $\delta_r$, we observe first, the formation of small clusters from individual particles and, second, their subsequent merging into larger clusters \cite{grant2011analyzing, whitelam2015statistical, grunwald2014patterns}. For $\delta_r=0.21\sigma$, there is no clear preference of a specific local structure, and thus, the particles have a probability to re-orient and bond with other clusters. This ultimately leads to quite compact clusters [see Fig.~\ref{Fig:3}(b)]. In contrast, for $\delta_r=0.1\sigma$ and $\delta_r=0.3\sigma$, the clusters have finer structures and are more spread out. This is a consequence of the stronger rigidity of bonds [also indicated by the slow decay of $B(\tau)$]. At the same time, the finer structures increase the likelihood of cluster merging, causing the value of $N_{max}/N$ to increase more rapidly than for $\delta_r=0.21\sigma$.

\subsubsection{Weakly non-reciprocal systems} \label{subsection: non-reciprocal system}
We now turn to equimolar binary mixtures characterized by finite values of $\kappa_{\text{nr}}$. In this section we mainly focus on weakly NR systems characterized by $\kappa_{\text{nr}}\leq0.3$. Exemplary simulation snapshots after $t=100\tau_B$ for the three anisotropy parameters $\delta_r$ are shown in Figs.~\ref{Fig:3}(d)-(f). In all cases the particles have aggregated into large, percolated clusters (note the periodic boundary conditions). This is in stark contrast to the corresponding reciprocal systems [see Figs.~\ref{Fig:3}(a)-(c)] where, at the same simulation time, multiple small clusters have formed.

These observations are substantiated by the behaviour of the quantity $N_{max}/N$ as function of time. Already for $\kappa_{\text{nr}}=0.1$, as seen from Fig.~\ref{Fig:4}(a), the size of the largest cluster increases much faster in the two-species systems than in their reciprocal counterparts. Moreover, $N_{max}/N$ reaches the value of 1 (corresponding to a single large cluster comprising all particles) before $50\tau_B$ for all anisotropy parameters, where the fastest increase occurs for the weakly anisotropic system ($\delta_r=0.1\sigma$). Still, full aggregation (i.e., $N_{max}/N \approx 1$) is even reached for the somewhat delicate case $\delta_r=0.21\sigma$ where, with reciprocal interactions, different cluster structures compete. 

\begin{figure*} [htp]
	\centering
	\includegraphics[width=1.0\textwidth]{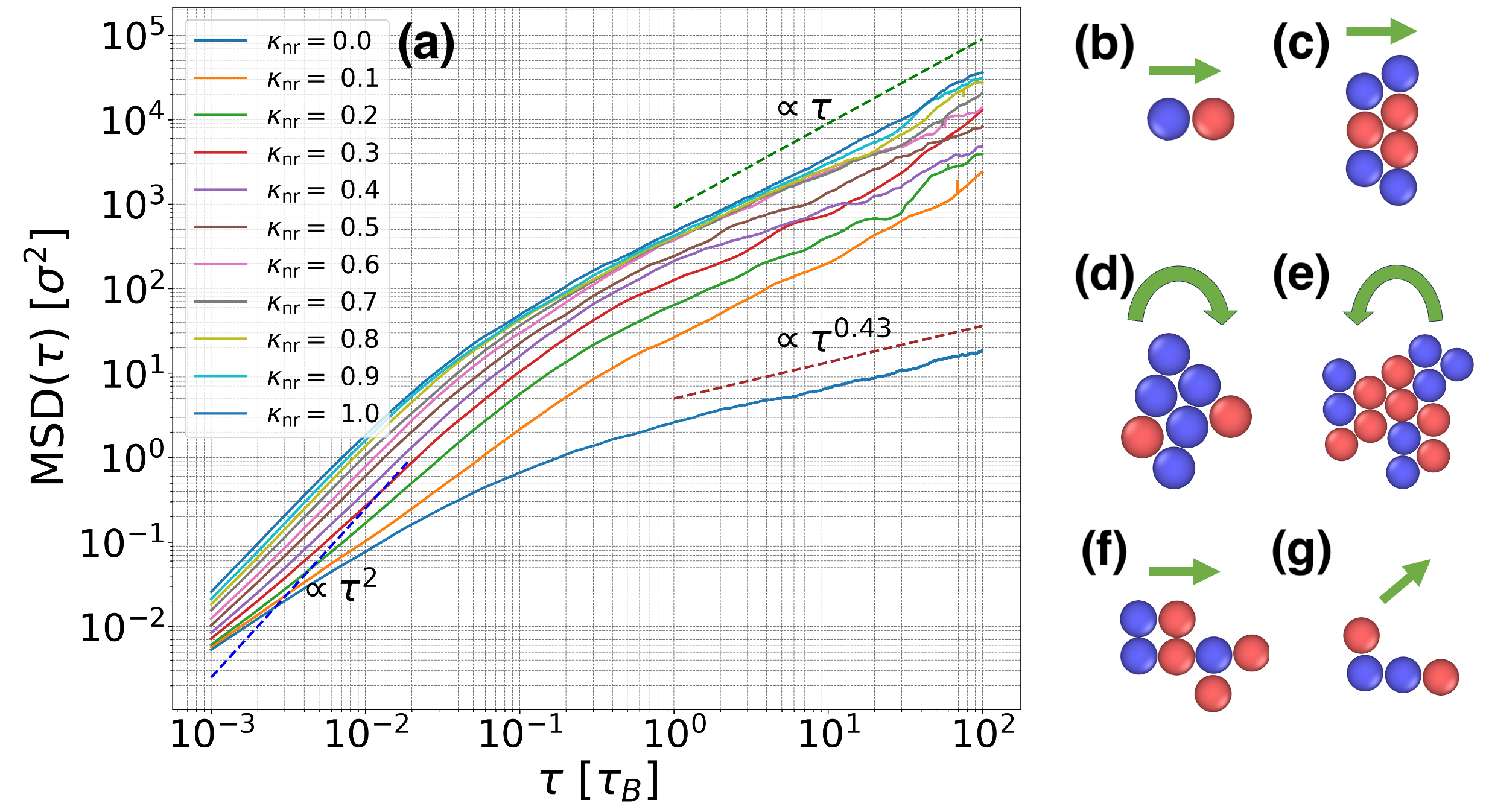}
	\caption{(a) MSD of the many particle system (averaged over all particle trajectories) with $\delta_r=0.1\sigma$ for the reciprocal case ($\kappa_{\text{nr}}=0$) and the NR case with different values of $\kappa_{\text{nr}}$. The coloured dashed lines indicate ballistic behaviour ($\text{MSD}(\tau)\propto\tau^2$, blue), diffusive behaviour ($\text{MSD}(\tau)\propto\tau^1$, green) and sub-diffusive behaviour ($\text{MSD}(\tau)\propto\tau^{0.43}$, brown). The data are obtained at $T^*=0.05$ and $\rho^*=0.3$. Examples of "active" colloidal molecules (b)-(e): $\delta_r=0.1\sigma$ and $\kappa_{\text{nr}}=0.1$, (f)-(g): $\delta_r=0.3\sigma$ and $\kappa_{\text{nr}}=0.1$. } \label{Fig:5}
\end{figure*}

\begin{center}
	\begin{table*} [htp]
		\caption{Velocity of "active" colloidal molecules} \label{Table:clus_vels}
		\begin{tabular}{| c | c | c |}
			\hline 
			"Active" molecule &  Timescale tracked & Velocity ($|\mathbf{v}|$ or $\omega$) \\ 
			\hline
			Fig.~\ref{Fig:5}(b) &  $t=100\tau_B$ &  $|\mathbf{\bar{v}}|=2.27\sigma/\tau_B$ \\  
			Fig.~\ref{Fig:5}(c) & $t=0.51 \tau_B$ & $|\mathbf{\bar{v}}|=1.18\sigma/\tau_B$ \\
			Fig.~\ref{Fig:5}(d) &  $t=4.05\tau_B$ &  $\bar{\omega}=-0.25 \text{rad}/\tau_B$ \\
			Fig.~\ref{Fig:5}(e) &  $t=6.27\tau_B$ &  $\bar{\omega}=0.34 \text{rad}/\tau_B$ \\
			Fig.~\ref{Fig:5}(f) &  $t=0.31\tau_B$ &  $|\mathbf{\bar{v}}|=1.70\sigma/\tau_B$ \\
			Fig.~\ref{Fig:5}(g) &  $t=0.14\tau_B$ &  $|\mathbf{\bar{v}}|=1.41\sigma/\tau_B$  \\
			\hline
		\end{tabular}
	\end{table*}
\end{center}

Another interesting observation for the particular case $\delta_r=0.21\sigma$ are the step-like jumps of $N_{max}/N$ even at long times. This reflects that clusters tend to break apart and re-form. Indeed, already in the corresponding reciprocal system, the aggregates are less stable than for other anisotropy parameters, as revealed by the bond correlation functions $B(\tau)$ plotted in Fig.~\ref{Fig:4}(b). These differences regarding the breaking of bonds persist for weakly NR systems ($\kappa_{\text{nr}}=0.1$), as seen from the functions $B(\tau)$ plotted in Fig.~\ref{Fig:4}(c). The data show that $B(\tau)$ decays more rapidly than in the reciprocal case for all three $\delta_r$ values, but the decay is most pronounced at $\delta_r=0.21\sigma$.
	
Given the NR induced effective “propulsion”, detected already for pairs of particles (see Sec.~\ref{section: nrp_propulsion}), we may speculate that propulsion also provides the main mechanism for the accelerated aggregation in our anisotropic many-particle system (indeed, similar observations have been made in simulation studies involving NR systems with isotropic interactions \cite{fehlinger2023collective, singh2017non, bartnick2016emerging}). 
	
To illustrate the emergence of effective propulsion in our many-particle system, we show in Fig.~\ref{Fig:5}(a) the system-averaged MSD defined as

\begin{equation}
	\text{MSD}(\tau)=\sum_{i=1}^{N}\left<\left[\mathbf{r}_i(t+\tau)-\mathbf{r}_i(t)\right]^2 \right>/N, \label{eq: MSD_nparticles}
\end{equation}

where the brackets $\left<..\right>$ denote a time as well as a noise average. The data in Fig.~\ref{Fig:5}(a) pertain to $\delta_r=0.1\sigma$ and different values of $\kappa_{\text{nr}}$. The general trends observed here hold for other values of $\delta_r$ as well. In the reciprocal case ($\kappa_{\text{nr}}=0$), the short-time MSD is nearly diffusive, but as time progresses, its slope decreases and we observe sub-diffusive behaviour with $\text{MSD}(\tau)\sim\tau^{0.43}$. That is, within the simulation time, the system does not reach the diffusive regime ($\text{MSD}(\tau)\sim\tau^1$) expected in a fully equilibrated many-particle system. We consider this as an indication of the aggregation into multiple, small clusters that barely change in time. This behaviour changes markedly already for $\kappa_{\text{nr}}=0.1$. Here, not only the initial increase of the MSD is faster (although not yet ballistic), but we also observe a diffusive regime at long times. As expected, the initial increase of the MSD becomes the faster the larger $\kappa_{\text{nr}}$ is, reflecting an increasing contribution of effectively propelled and, thus, ballistically moving particles. Moreover, all NR systems reach the diffusive limit.

So far we have considered system-averaged quantities. Further interesting information is gained when we analyze the simulations on a particle level. In particular, to better understand the early stages of aggregation, we have searched for the type of aggregated structures spontaneously formed at short times (i.e., before the merging into large, percolated clusters). Indeed we have found a variety of initial structures (or “molecules”) that, in the NR case and depending on the parameters, can propel and even rotate. The formation of such small colloidal molecules has already been reported in recent literature \cite{niu2018dynamics, schmidt2019light, varma2018clustering, gonzalez2019directed, soto2014self, grauer2021active, hauke2020clustering, bartnick2016emerging} on isotropic systems. In the present systems, anisotropy adds an additional layer of complexity. Examples (for $\kappa_{\text{nr}}=0.1$) are shown in Figs.~\ref{Fig:5}(b)-(g). The green arrows next to the molecules indicate their respective directions of propulsion or rotation. The smallest building block are the pairs of opposite species ("active colloidal dimers") already discussed in Sec.~\ref{section: nrp_propulsion}. These pairs then combine with other pairs or single particles. The features of the resulting larger “molecules” depend on the anisotropy parameter $\delta_r$, since increasing anisotropy causes the active dimers to be more and more aligned along the direction of induced dipoles (i.e., the $x-$ or $y-$ axis). This alignment can tune the composition and orientation of the formed “molecules”. Therefore, at weak anisotropy ($\delta_r=0.1\sigma$) we often find rather compact structures that, moreover, can rotate. In contrast, at larger anisotropy, the molecules consist of particles aligned strictly along the axes, and rotations do not occur. In order to quantify the motion of these "active" colloidal molecules, we track the sample molecules shown in Figs.~\ref{Fig:5}(b)-(g) during the timescale within which they remain stable in the simulation (i.e., they do not deform or merge into another cluster) and calculate their average centre-of-mass translational ($|\mathbf{\bar{v}}|$) and rotational ($\bar{\omega}$) velocities. Details about the calculation of the velocities are provided in the Appendix \ref{appendix: clus_vels}. The results are summarised in Table~\ref{Table:clus_vels}. Compared to the centre-of-mass translational velocities $|\mathbf{\bar{v}}|$ of the larger aggregates in Figs.~\ref{Fig:5}(c), (f) and (g), the pair of opposite species in Fig.~\ref{Fig:5}(b) was found to have a larger $|\mathbf{\bar{v}}|$. This is due to the fact that the effective centre-of-mass velocity (both translational and rotational) of the larger aggregate depends on the orientations of the self-propelling pairs within them. This observation is in agreement with other studies where the centre-of-mass velocities of "active" colloidal molecules were investigated \cite{niu2018dynamics, schmidt2019light, varma2018clustering, gonzalez2019directed, soto2014self, grauer2021active, hauke2020clustering}. The aggregates that we study here consist of configurations where the orientations were able to generate a propulsion mechanism, albeit being lower in magnitude than the basic building block, that is, a pair of particles of different species.

\begin{figure} [htp]
	\centering
	\includegraphics[width=0.5\textwidth]{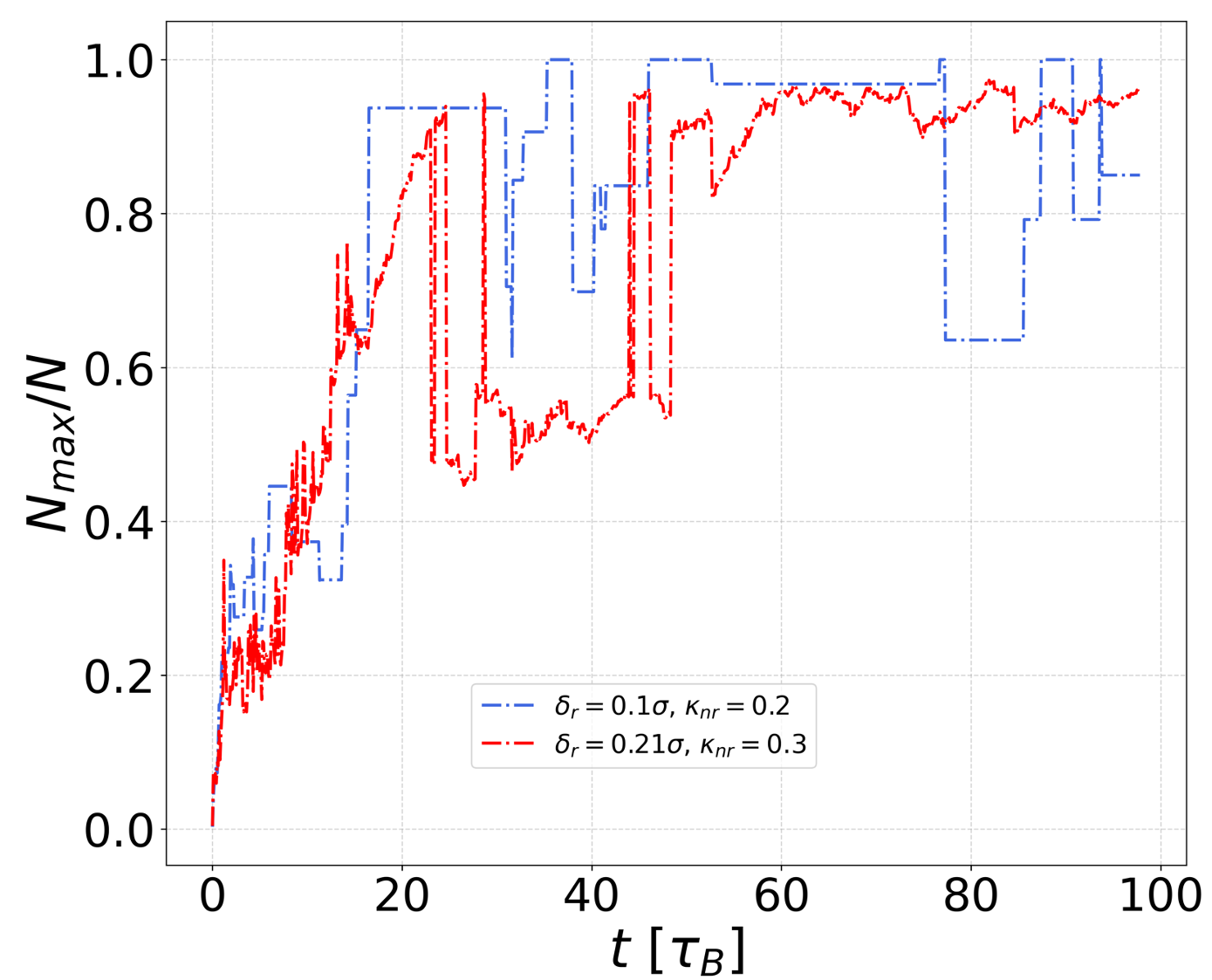}
	\caption{Largest cluster size ($N_{max}/N$) as a function of time for $\delta_r=0.1\sigma$ and $\kappa_{\text{nr}}=0.2$ (dashed blue lines), and $\delta_r=0.21\sigma$ and $\kappa_{\text{nr}}=0.3$ (dashed red lines).} \label{Fig: breakups}
\end{figure}

As we increase $\kappa_{\text{nr}}$, the effective velocity of mixed pairs of A and B particles increases as well [see Fig.~\ref{Fig:2}(c)], thereby speeding up the active “molecules” that form from these self-propelled particle pairs. As an accompanying effect, we find that the large clusters formed at later times of aggregation break up and recombine more frequently. These trends are revealed when we inspect again, in Fig.~\ref{Fig: breakups} the size of the largest cluster ($N_{max}/N$) as a function of time, this time at somewhat larger values of $\kappa_{\text{nr}}$ than those considered in Fig.~\ref{Fig:4}. As shown for $\delta_r=0.1\sigma$ and $\kappa_{\text{nr}}=0.2$, the size of the largest cluster grows and drops in large steps, reflecting that clusters split up and re-merge. However, Increasing $\kappa_{\text{nr}}$ even further causes the induced propulsion to start counteracting the clustering mechanism. At this stage, some of the self-propelled particle pairs can escape from large clusters. An example pertaining to $\delta_r=0.21\sigma$ and $\kappa_{\text{nr}}=0.3$ is also plotted in Fig.~\ref{Fig: breakups}. Here, in addition to the step-like changes of $N_{max}/N$, we observe pronounced fluctuations that can be attributed to the escape of particle pairs from the largest cluster. We will come back to this cluster destruction in the next Section~\ref{subsection: phase separation and disorder}.

\subsection{Phase separation and disorder in strongly non-reciprocal systems}   \label{subsection: phase separation and disorder}

\begin{figure*} [htp!]
	\centering
	\includegraphics[width=0.95\textwidth]{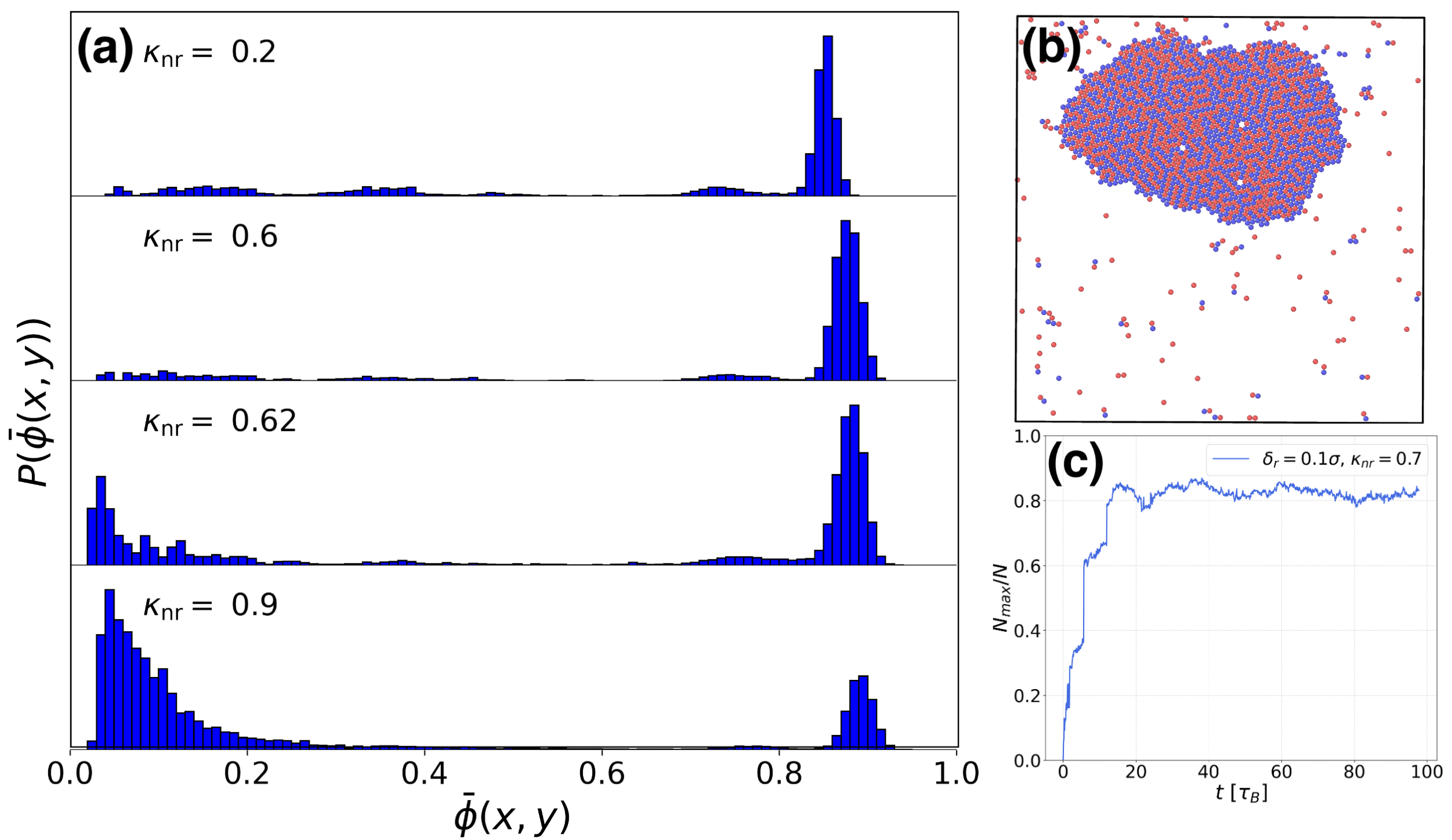}
	\caption{(a) Distribution of the position resolved local-area fractions $P(\bar{\phi}(x,y))$ at different values of $\kappa_{\text{nr}}$ for the $\delta_r=0.1\sigma$ case. (b) Snapshot of a simulation at $\delta_r=0.1\sigma$ and $\kappa_{\text{nr}}=0.7$ started from a random, uniform distribution of particles at $\rho^*=0.3$ at $T^*=0.05$ after a simulation time of $t=100\tau_B$. The blue particles correspond to particles of species A and the red ones to species B. (c) Largest cluster size ($N_{max}/N$) as a function of time.} \label{Fig:6}
\end{figure*}

Upon further increase of the NR parameter $\kappa_{\text{nr}}$, we leave the parameter regime of cluster aggregation (at the density and temperature considered). Indeed we observe, depending on the degree of anisotropy measured by $\delta_r$, essentially two types of behaviours: phase separation into a (mixed) dense and (mixed) dilute phase ($\delta_r=0.1\sigma$), or a completely disordered state ($\delta_r=0.21\sigma$ and $\delta_r=0.3\sigma$).

We first discuss the phase separation for weakly anisotropic systems ($\delta_r=0.1\sigma$). As an illustration, Fig.~\ref{Fig:6}(b) shows a simulation snapshot at $\kappa_{\text{nr}}=0.7$ after a simulation time of $t=100\tau_B$, started from a mixed, random distribution of particles. The snapshot clearly reveals the co-existence of a dense, large cluster of particles of both species with a dilute region consisting of many single particles and particle pairs. The presence of phase separation is also reflected in the (time-averaged) distribution of the local area fraction, $P(\bar{\phi}(x,y))$, plotted in Fig.~\ref{Fig:6}(a) for different values of $\kappa_{\text{nr}}$ for the $\delta_r=0.1\sigma$ case (for details of the calculation, see Appendix \ref{appendix: position resolved local area fracs}). A phase-separated state is indicated by a double-peak structure of $P(\bar{\phi}(x,y))$ \cite{blaschke2016phase, liao2018clustering}, which is clearly visible here for $\kappa_{\text{nr}}\geq0.62$. The location of the two peaks then indicate the mean local area fractions of the "coexisting" states. Interestingly, such a double-peak structure does not occur for the two larger values of $\delta_r$ discussed below, indicating that (too) strong anisotropy of the field-induced interactions destroys phase separation. Within the phase-separated state observed at $\delta_r=0.1\sigma$, single particles and particle pairs keep leaving and re-entering the cluster, revealing a dynamically balanced situation. This is seen from the quantity $N_{max}/N$ as a function of time plotted in Fig.~\ref{Fig:6}(c). The size of the largest cluster exhibits persistent, small fluctuations around a mean value of around 0.8 measuring the cluster's average size.

\begin{figure*} [htp]
	\centering
	\includegraphics[width=0.95\textwidth]{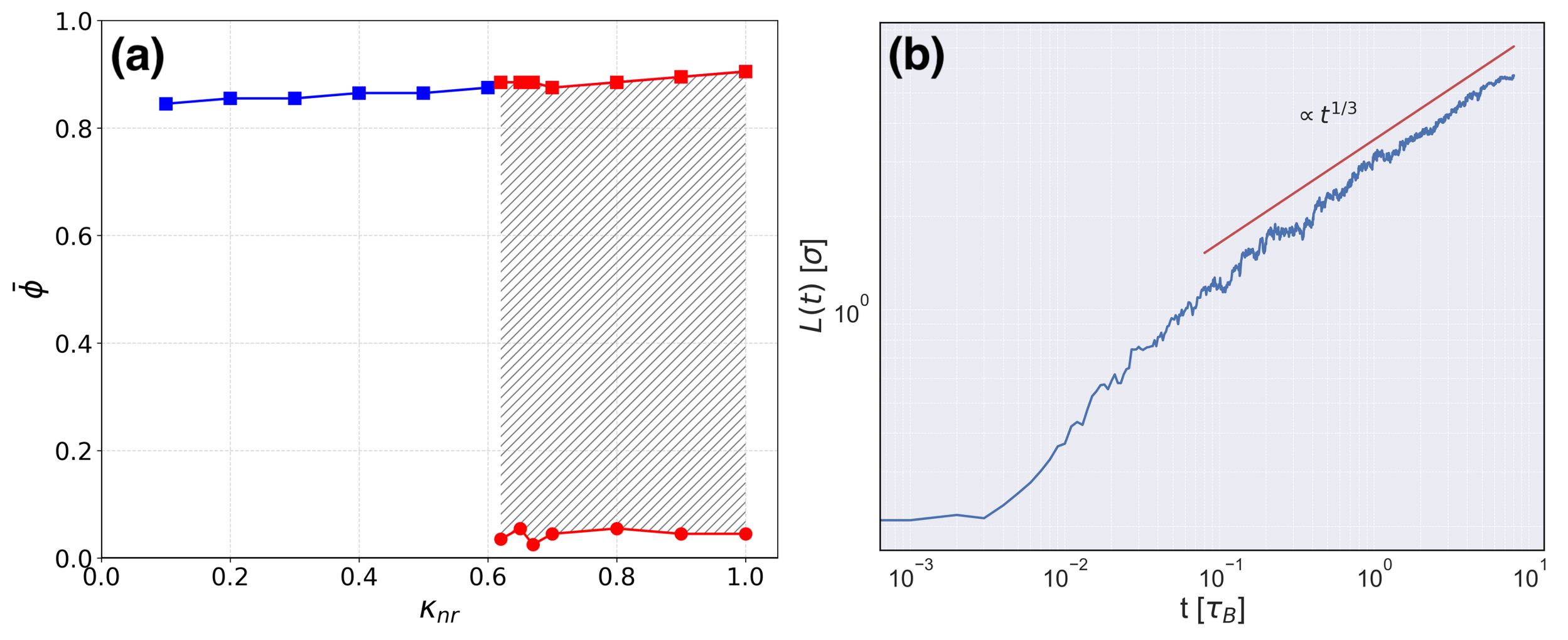}
	\caption{(a) Local area fractions of the observed phases (only dense phase at $\kappa_{\text{nr}}<0.62$ and co-existing dense and dilute phases at $\kappa_{\text{nr}}\geq0.62$) as a function of the parameter $\kappa_{\text{nr}}$ for the $\delta_r=0.1\sigma$ case. The squares represent the dense phase and the circles represent the dilute phase. The colour blue is used to represent the data points corresponding to the single phase system and the colour red for the data points corresponding to the phase co-existing system. The co-existence region between the binodal curves is hatched for visualization. (b) Characteristic domain length $L(t)$ as a function of time $t$. The red line indicates the fitted line corresponding to the scaling exponent $\alpha$ of $L(t) \propto t^{\alpha}$ with $\alpha=1/3$. } \label{Fig: phase_sep}
\end{figure*}

In order to characterize the phase separation in the $\delta_r=0.1\sigma$ case, we now construct the binodal using the distributions of the local area fractions at different values of $\kappa_{\text{nr}}$ [Fig.~\ref{Fig:6}(a)]. As a first step, we determine the area fractions of the respective phases (dense phase at $\kappa_{\text{nr}}<0.62$ and co-existing dense and dilute phases at $\kappa_{\text{nr}}\geq0.62$) by identifying the position of the peaks. The resulting area fractions of the two phases as a function of $\kappa_{\text{nr}}$ for the $\delta_r=0.1\sigma$ case are shown in Fig.~\ref{Fig: phase_sep}(a). We note that the plot has some similarity to the binodal curves reported for motility-induced phase separation of active Brownian particles \cite{redner2013structure}. Here, we see that the system phase separates as $\kappa_{\text{nr}}$ increases. 

We have also considered the phase separation kinetics by calculating the exponent $\alpha$ associated with the growth of characteristic domain length scale $L(t)$ in time (for details regarding the calculation of $L(t)$ see Appendix \ref{appendix: length_scale}). We calculate $L(t)$ for the $\delta_r=0.1\sigma$ and $\kappa_{\text{nr}}=0.8$ case from 50 different noise realisations of $100\tau_B$ each. We note that, since we perform simulations of a system consisting of $N=1800$ particles, finite size effects can cause the value of $L(t)$ to saturate after a certain time. Hence, for calculating $\alpha$ we use data for the time during which $L(t)$ was observed to grow. In Fig.~\ref{Fig: phase_sep}(b), we show $L(t)$ as a function of time. Fitting the data in Fig.~\ref{Fig: phase_sep}(b) to the power-law relation $L(t)\propto t^{\alpha}$, we obtain $\alpha\simeq0.33=1/3$. The value of $\alpha=1/3$ is in agreement with the expected scaling factor for systems showing phase separation with diffusive transport of the order parameter \cite{stenhammar2013continuum, chaikin1995principles, bray2002theory}.

The observation of phase separation induced by NR conforms with results from previous simulation studies of NR mixtures with isotropic interactions (see, e.g., Ref.\cite{agudo2019active}). For such systems, it was also shown that the character of phase separation (e.g. condensation versus de-mixing) depends on the overall volume fraction and mixing ratio \cite{agudo2019active}. We expect that variation of these parameters (which were kept constant here) would also lead to different types of phase separation in the present, weakly anisotropic mixture.

\begin{figure*} [htp]
	\centering
	\includegraphics[width=0.95\textwidth]{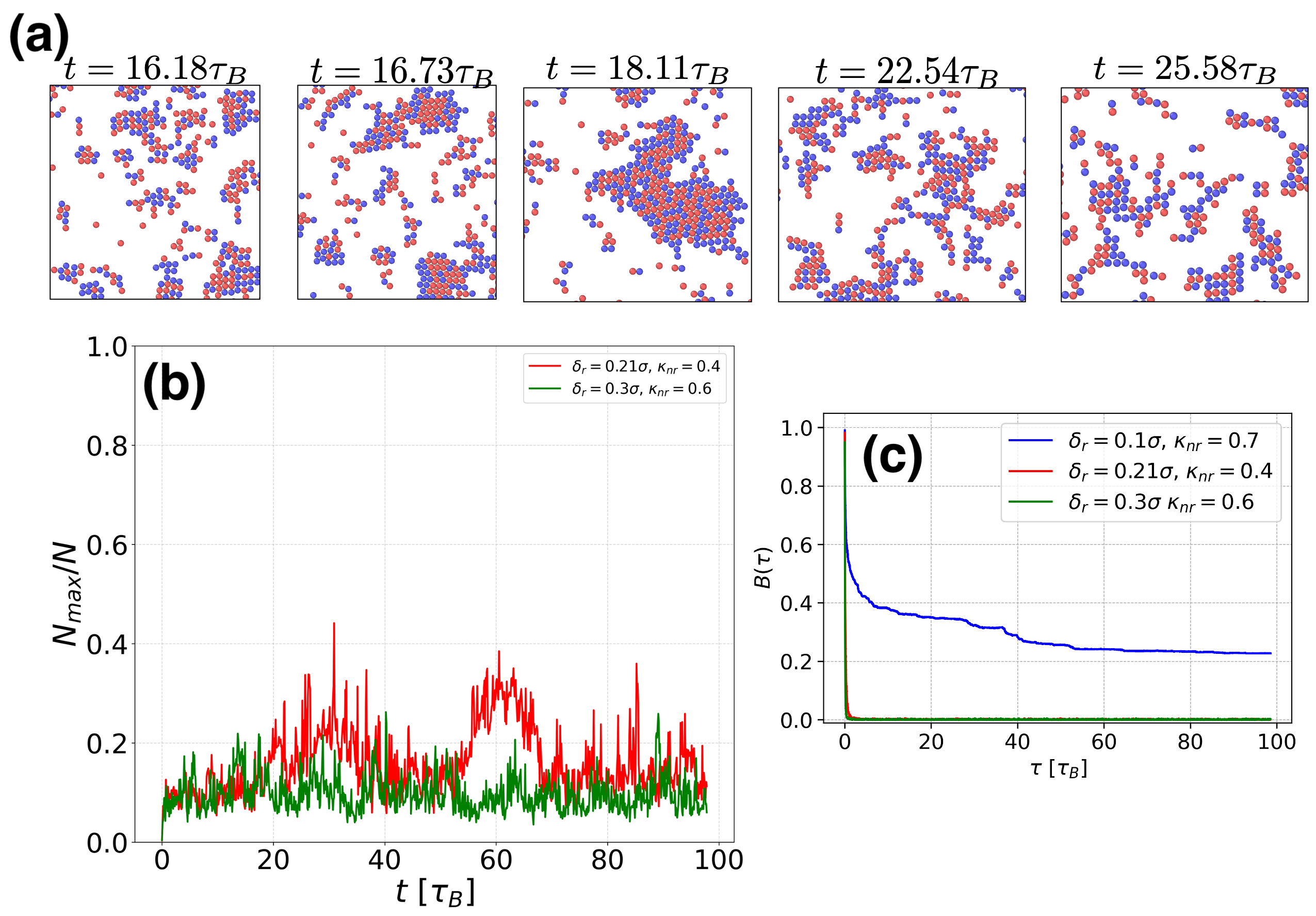}
	\caption{(a) Zoomed-in snapshots at $\delta_r=0.21\sigma$, $\kappa_{\text{nr}}=0.4$, and different times, showing the formation and the subsequent dissolution of a small cluster. (b) Largest cluster size ($N_{max}/N$) as a function of time for $\delta_r=0.21\sigma$, $\kappa_{\text{nr}}=0.4$ and $\delta_r=0.3\sigma$, $\kappa_{\text{nr}}=0.6$. (c) Bond auto-correlation function for $\delta_r=0.1\sigma$, $\kappa_{\text{nr}}=0.7$; $\delta_r=0.21\sigma$, $\kappa_{\text{nr}}=0.4$; and $\delta_r=0.3\sigma$, $\kappa_{\text{nr}}=0.6$.} \label{Fig:7}
\end{figure*}

We now turn to the behaviour of systems with stronger anisotropy ($\delta_r=0.21\sigma$ and $0.3\sigma$) and NR. Here, there is no large-scale phase separation, rather the systems develop a disordered, yet highly dynamical state. Exemplary snapshots obtained at $\delta_r=0.21\sigma$ and $\kappa_{\text{nr}}=0.4$ are presented in Fig.~\ref{Fig:7}(a), revealing the formation and subsequent breaking of a small cluster as time proceeds. Corresponding results for $N_{max}/N(t)$ are plotted in Fig.~\ref{Fig:7}(b), along with data for another, comparable parameter combination. In both cases, $N_{max}/N$ exhibits rapid fluctuations around a rather low average value in the range 0.1-0.2, which is substantially smaller than in the case of phase separation [see Fig.~\ref{Fig:6}(c)] or in weakly NR systems [see Fig.~\ref{Fig:4}(a)]. Similar signs of disorder are seen at even higher values of the NR parameter.

To shed some light on the impact of anisotropy at large $\kappa_{\text{nr}}$ from a microscopic point of view, we investigate the fate of the small aggregates formed at initial stages of aggregation. As discussed in Section~\ref{subsection: non-reciprocal system}, NR leads to the formation of self-propelled colloidal “molecules” (see Fig.~\ref{Fig:5}) at early times that turn out to be quite stable for small values of $\kappa_{\text{nr}}$. However, upon increase of $\kappa_{\text{nr}}$, the induced propulsion starts to counteract the aggregation; in other words, the small molecules quickly fall apart due to the large propulsion velocities of the particle pairs. This can be seen from the bond-autocorrelation functions plotted in Fig.~\ref{Fig:7}(c) for all three values of $\delta_r$ and representative NR parameters $\kappa_{\text{nr}}\geq0.4$. In all cases, the correlation function decays much faster in time than at low NR [see Fig.~\ref{Fig:4}(c)], confirming our previous statement of the reduced stability of “molecules” (in fact, an appreciable number of bonds at longer times is found only for $\delta_r=0.1\sigma$). Thus, at large $\kappa_{\text{nr}}$, the system's structure is rather dominated by single particles and self-propelled A-B pairs. If these pairs could freely move in any direction, one would expect a behaviour similar to that of repulsive active particle systems where, the combination of propulsion and repulsion indeed leads to trapping and eventually to “motility-induced” phase separation \cite{buttinoni2013dynamical}. Such a (essentially) free motion indeed occurs in our system when the anisotropy parameter is small ($\delta_r=0.1\sigma$). We thus speculate that trapping of pairs (that can approach each other from any direction), together with the angle-averaged attraction via the anisotropic potential, is essentially the mechanism behind the observed phase separation at weak anisotropy (see Fig.~\ref{Fig:6}). The situation is different at $\delta_r=0.21\sigma$ and $\delta_r=0.3\sigma$ where the anisotropic interactions force the particles to form pairs only along the $x-$ and $y-$ axis. These axes also restrict the direction of motion of the pairs. In this situation, trapping can only occur in the (rather unlikely) case that oppositely moving particle pairs have a head-on collision. Even if larger clusters form, they will be easily disturbed by other, freely moving particles or particle pairs and thus quickly dissolve [see Fig.~\ref{Fig:7}(a)]. Thereby, large interaction anisotropy can suppress phase separation and lead to a disordered state.

\begin{center}
	\begin{table*} [htp]
		\caption{Characteristics of the different states} \label{Table:1}
		\begin{tabular}{| c | c | c |}
			\hline 
			State &  $N_{max}/N$ & No. of peaks in $P(\bar{\phi}(x,y))$ \\ 
			\hline
			Aggregates &  $\geq0.7$ with large, step-like fluctuations & one (at large $\bar{\phi}$) \\  
			Phase separation & $\sim0.7-0.9$ with small fluctuations & two (at low and large $\bar{\phi}$)\\
			Disorder &  $\leq0.03$ with large fluctuations & one (at low $\bar{\phi}$) \\
			\hline
		\end{tabular}
	\end{table*}
\end{center}

\begin{figure} [htp]
	\centering
	\includegraphics[width=0.5\textwidth]{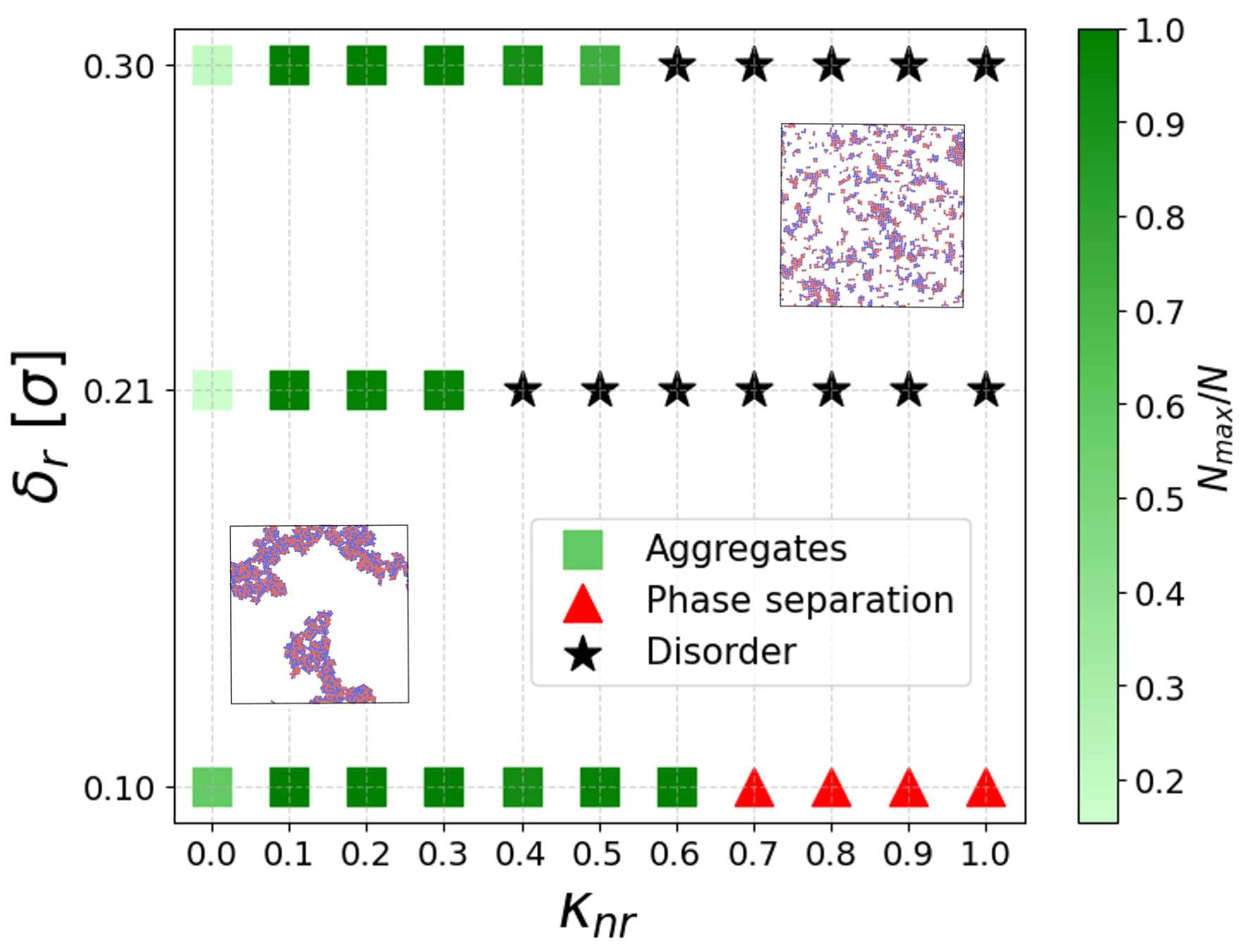}
	\caption{State diagram after $t=100\tau_B$ at $\rho^*=0.3$ and $T^*=0.05$ in the plane spanned by the NR parameter $\kappa_{\text{nr}}$ (horizontal axis) and the anisotropy parameter $\delta_r$ (vertical axis).} \label{Fig:8}
\end{figure}

As an overview of the behaviours seen in the binary mixture at different degrees of NR and anisotropy, we present in Fig.~\ref{Fig:8} a state diagram in the parameter plane spanned by $\kappa_{\text{nr}}$ and $\delta_r$. We have identified three overall "states" whose characteristics are summarize in Table \ref{Table:1}. For weak NR ($\kappa_{\text{nr}}\leq0.3$), the systems behave similar to the reciprocal reference system insofar that there is no large scale phase separation, but rather aggregation into clusters. In this regime, the main effect of NR is that the clusters, at a given time, are much larger and “thermalized” (i.e., particle can escape and re-enter), in contrast to the rigid structures formed in the reciprocal case. In this sense, our results show that NR shortcuts the overall aggregation process. On the particle level, we find that weak NR leads to the formation of active “molecules” whose character and stability depends on the anisotropy of interactions.

At larger degrees of NR, the overall behaviour strongly depends on $\delta_r$. In particular, phase separation is only found for weakly anisotropic systems (dominated by freely moving self-propelled pairs) whereas strongly anisotropic interactions lead to disordered configurations where only small clusters occasionally occur.

\section{Conclusion}  \label{section: conclusion}
This study has been devoted to the impact of NR on a self-aggregating system
of passive colloidal particles with anisotropic interactions induced by external fields. To this end, we have considered a binary version of a model introduced earlier \cite{kogler2015generic}, where NR occurs as a prefactor of the anisotropic inter-species interactions. We then performed extensive BD simulation for a range of parameters regulating the degree of NR, on the one hand, and anisotropy, on the other hand.

Consistent with earlier studies involving isotropic interactions \cite{fehlinger2023collective, soto2014self, soto2015self, ai2023brownian, varma2018clustering, schmidt2019light, grauer2021active, ruckner2007chemically, valadares2010catalytic, reigh2015catalytic} we found that NR can lead to the formation of effectively self-propelled pairs (or larger “molecules”) of particles of different species. In this regime, the degree of anisotropy mainly determines the molecules’ structure and modes of motion. The NR-induced motion is a signature that the system is indeed driven out of equilibrium (as one would generally expect in an NR system). For relatively weak NR, the effective motility enhances the aggregation processes in the sense that, at the same simulation times, the size of clusters is substantially larger than in the reciprocal case, and the bonds are less stable. This overall “annealing” effect is essentially independent of the degree of anisotropy. We note, in this context, an interesting analogy to other types of non-equilibrium self-assembly: for example, in dense systems of triblock Janus colloids that tend to stuck in long-lived transients \cite{mallory2019activity, trubiano2021thermodynamic}, introducing intrinsic activity to the particles was shown to overcome free energy barriers, thereby allowing the system to form faster the thermodynamically favoured (Kagome lattice) state \cite{mallory2019activity}. In the present system, the role of activity is, to some extent, replaced by NR interactions. However, when the anisotropic interactions are strongly NR, we find different behaviour where the clusters become unstable: in this regime, weakly anisotropic systems exhibit phase separation, whereas strong anisotropy leads to complete disorder. 

In this study we have considered the collective behaviour only at one density, composition and temperature. It seems likely that by varying these parameters the phase boundaries in Fig.~\ref{Fig:8} are shifted, or even new states emerge (as reported in the case of isotropically interacting particles \cite{agudo2019active, chiu2023phase}). Another open question is what would happen in the case of antagonistic couplings where the interactions are not only NR, but have even different signs. Judging from recent studies on other NR systems \cite{saha2020scalar, mandal2022robustness, you2020nonreciprocity, fruchart2021non, kreienkamp2022clustering, agudo2019active, chiu2023phase}, such antagonistic couplings could open up the intriguing possibility of observing spontaneously formed time-dependent states, e.g. travelling patterns or chiral motion, whose interplay with aggregation processes has yet to be explored. Work in these directions is under way.

\begin{acknowledgements}
This work was funded by the Deutsche Forschungsgemeinschaft (DFG, German Research Foundation), project number 449485571. 
\end{acknowledgements}

\section*{Author declarations}
\subsection*{Conflict of Interest}
The authors have no conflicts to disclose. 

\subsection*{Author contributions}
\textbf{Salman Fariz Navas:} Formal analysis (lead); software (lead); writing - original draft preparation (lead). \textbf{Sabine H. L. Klapp}: Conceptualization (lead); writing – review and editing (lead), funding acquisition (lead).

\section*{Data Availability Statement}
The data that support the findings of this study are available from the corresponding authors upon reasonable request.

\appendix
\section{Threshold value of $\delta_r$}  \label{appendix: delr_threshold}
It has been shown in Ref. \cite{kogler2015generic} that the nature of aggregates in the reciprocal system depends on the parameter $\delta_r$. The threshold value of $\delta_r$, where the aggregates change from being predominantly hexagonally ordered to quadratic ordering is found to be at $\delta_r=0.21\sigma$ \cite{kogler2015generic}. At $\delta_r=0.21\sigma$,  neither the hexagonal structure, nor the quadratic structure is strongly preferred leading to a competition between the two structures. In order to estimate the threshold value, two order parameters, namely the mean co-ordination number ($\bar{z}$) and the mean orientational order parameters ($\bar{\Phi}_4$ and $\bar{\Phi}_6$) were used \cite{kogler2015generic}. The order parameters are calculated at a time point after the aggregates become quasi-stationary ($t=100\tau_B$) for values of $\delta_r$ from $0.05-0.35$ $\sigma$. The co-ordination number $z_k$ is defined as the number of nearest neighbours corresponding to each particle $k$ at time $t$. The mean co-ordination number is then defined as
\begin{equation}
	\bar{z}=\frac{1}{N}\sum_{k=1}^{N}z_k.
\end{equation}

\begin{figure*} [htp]
	\centering
	\includegraphics[width=0.90\textwidth]{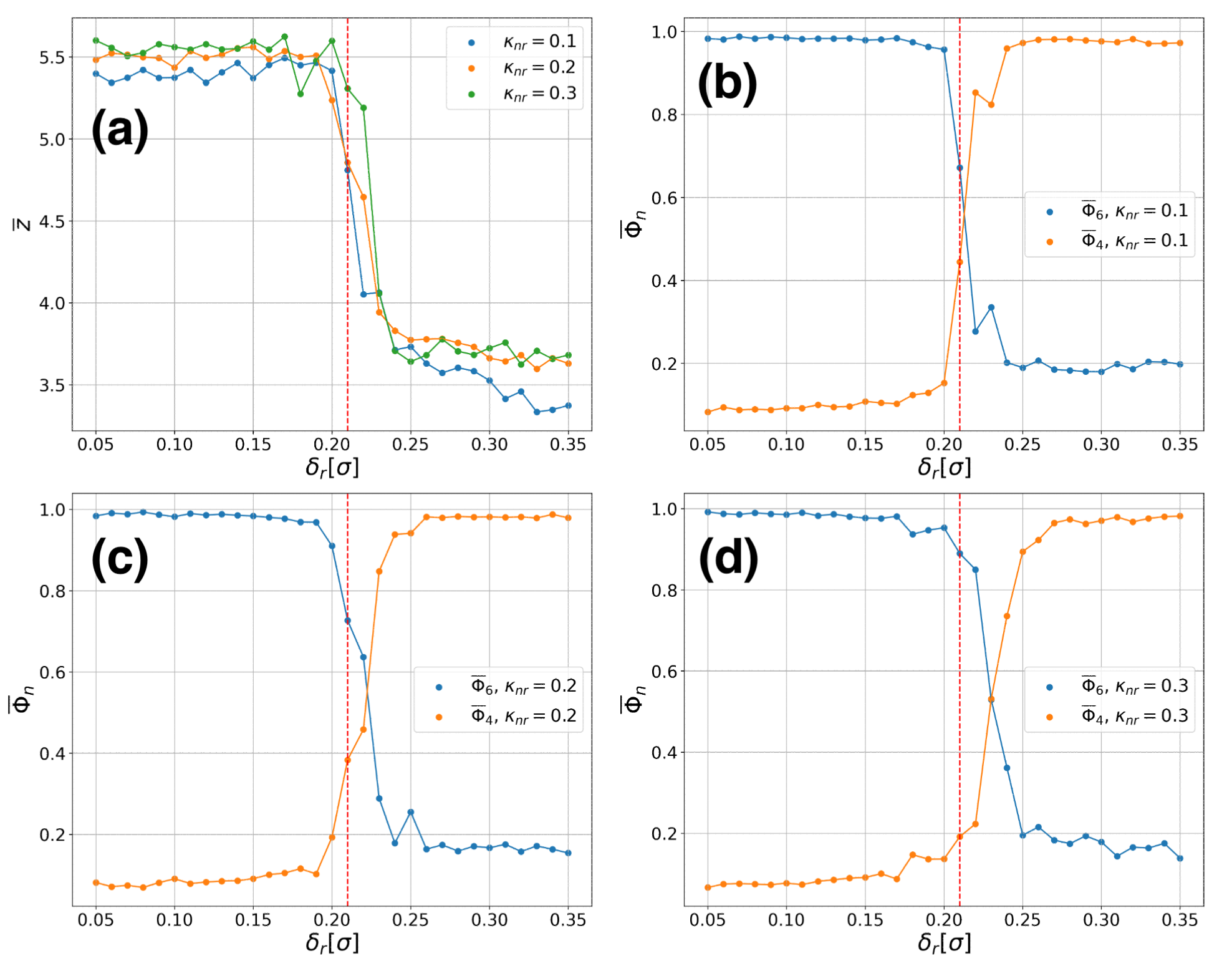}
	\caption{(a) Mean co-ordination number $\bar{z}$ as a function of $\delta_r$ for non-reciprocal systems with $\kappa_{\text{nr}}=0.1$, $0.2$ and $0.3$. Mean orientational order parameters $\bar{\Phi}_4$ and $\bar{\Phi}_6$ as a  function of $\delta_r$ for the non-reciprocal system with $\kappa_{\text{nr}}=$ (b) $0.1$, (c) $0.2$ and (d) $0.3$. The order parameters are calculated at $t=100\tau_B$. The vertical dashed line indicates the threshold value observed in the reciprocal system ($\delta_r=0.21\sigma$). } \label{Fig:appendix}
\end{figure*}

The orientational order parameters gives information about the orientation of the $z_k$ nearest neighbour particles around the central particle $k$ at time $t$ and is defined as
\begin{equation}
	\Phi_n^k=\frac{1}{z_k}\left|\sum_{l=1}^{z_k}e^{in\theta_{kl}} \right|,
\end{equation} 

where $\theta_{kl}$ is the angle between the vector $\mathbf{r}_{kl}=\mathbf{r}_k-\mathbf{r}_l$ connecting the central particle under consideration (denoted by index $k$) and one of its neighbours (denoted by index $l$), and the $x-$axis. The value of $\Phi_n^k$ is equal to 1 if the order is perfect in the sense that the particle $k$ has exactly $n$ neighbours and each each pair $(k,l)$ forms an angle of $\theta_{kl}=2\pi/n$ with the adjacent pair. For our system, we consider the orientational order parameters $\Phi^k_4$ (indicating quadratic ordering) and $\Phi^k_6$ (indicating hexagonal ordering). For systems with thermal fluctuations, a particle belonging to a quadratic structure is expected to have a value of $\Phi^k_4$ close to 1 and $\Phi^k_6$ close to 0 and vice-versa for a particle belonging to a hexagonal structure. The mean orientational order parameter of the system is then given by,
\begin{equation}
	\bar{\Phi}_n=\frac{1}{N}\sum_{k=1}^{N}\Phi_n^k. \label{eq:mean_abop}
\end{equation}

We plot these order parameters as a function of $\delta_r$ in Fig.~\ref{Fig:appendix}. The threshold can be identified as the value of $\delta_r$ at which the order parameter indicating a specific ordering (either hexagonal or quadratic) increases sharply while the one indicating the other decreases. This was found to occur at $\delta_r=0.21\sigma$ for the reciprocal system ($\kappa_{\text{nr}}=0$) \cite{kogler2015generic}. In Fig.~\ref{Fig:appendix}, we show data for various non-reciprocal systems, specifically the mean co-ordination number ($\bar{z}$) and the mean orientational order parameters ($\bar{\Phi}_4$ and $\bar{\Phi}_6$) as a function of $\delta_r$. We consider the values of $\kappa_{\text{nr}}$ where aggregation was observed to occur for all the values of $\delta_r$ considered i.e., $\kappa_{\text{nr}}=0.1$, $0.2$ and $0.3$. In Fig.~\ref{Fig:appendix}(b) it can be seen that the threshold still occurs at $\delta_r=0.21\sigma$ for $\kappa_{\text{nr}}=0.1$. The order parameter $\bar{\Phi}_6$ is seen to sharply decrease from $\bar{\Phi}_6=0.96$ at $\delta_r=0.20\sigma$ to $\bar{\Phi}_6=0.67$ at $\delta_r=0.21\sigma$, whereas $\bar{\Phi}_4$ increases from $\bar{\Phi}_4=0.15$ to $\bar{\Phi}_4=0.44$. However, such a sharp transition is not observed for $\kappa_{\text{nr}}=0.2$ and $0.3$ where the transition becomes more gradual [see Figs.~\ref{Fig:appendix}(c) and (d)]. The threshold value of $\delta_r$ also appears to shift towards slightly higher values in these two cases with $\kappa_{\text{nr}}>0$.

\section{Normalization of $U_{\text{DIP}}(\mathbf{r}_{ij})$} \label{appendix: U_dip_norm}
To allow for the comparison of $U_{\text{DIP}}(\mathbf{r}_{ij})$ at different $\delta_r$, we normalize $U_{\text{DIP}}(\mathbf{r}_{ij})$ such that it has the same value at the minima for different $\delta_r$, i.e.,
\begin{equation}
	\tilde{U}_{\text{DIP}}(\mathbf{r}_{ij})=U_{\text{DIP}}(\mathbf{r}_{ij})\frac{u}{U_{\text{DIP}}(\sigma \mathbf{\hat{e}}_{\alpha})}. \label{eq:U_DIP_norm}
\end{equation}

Here, $u=-0.2804\epsilon$  is a constant (chosen consistently with Ref. \cite{kogler2015generic}) calculated from the un-normalized potential $U_{\text{DIP}}(\sigma \mathbf{\hat{e}}_{\alpha})$ for $\delta_r=0.3\sigma$.  

\section{Extracting the effective propulsion velocity $v_0$} \label{appendix: v_0_fit}
The MSD of a single particle is defined as $\text{MSD}(\tau)=\left<(\mathbf{r}(t+\tau)-\mathbf{r}(t))^2 \right>$. The analytic expression for the MSD of an active Brownian particle (ABP) is given by\cite{howse2007self, ten2011brownian, zottl2016emergent}, 
\begin{equation}
	\text{MSD}(\tau) = 4D_t\tau + 2v_0^2 \tau_r^2\left( \frac{\tau}{\tau_r}+ \exp \left(-\frac{\tau}{\tau_r}\right)-1 \right). \label{eq: MSD_ABP}
\end{equation}
where $D_t=k_BT/\gamma$ is the translational diffusion coefficient and $\tau_r$ is the persistence time of the ABP. At short times, when $\tau\ll\tau_r$, Eq.~(\ref{eq: MSD_ABP}) can be re-written as
\begin{equation}
	\text{MSD}(\tau)=4D_t\tau+v_0^2\tau^2,  \label{eq: MSD_ABP_ballistic}
\end{equation}
indicating a combination of diffusive motion and ballistic motion with velocity $v_0$. In the present work, we use Eq.~(\ref{eq: MSD_ABP_ballistic}) to fit our data of the MSD of the non-reciprocally interacting particle pair. We thereby extract an effective propulsion velocity corresponding to $v_0$. Since the MSD is observed to scale ballistically throughout the length of the simulation [see Fig.~\ref{Fig:2}(a)], we use the entire range of data for performing the fit. 

\section{Bond auto-correlation function} \label{appendix: BAC}
Bond auto-correlation functions are dynamical quantities that can be used to study the lifetime of bonds that form between particles in self-aggregating systems. To calculate the bond auto-correlation function in time ($B(\tau)$) for our system, we first define an $N\times N$ matrix $\mathbf{b}(t)$ whose elements $b_{ij}(t_0)$ are assigned a value of 1 if particles $i$ and $j$ are bonded and 0 if they are not, at time $t$. Two particles are considered to be mutually bonded if their centre-to-centre distance is smaller than $r_{cl}=1.15\sigma$ \cite{kogler2015generic}. The bond auto-correlation function is then defined as \cite{kogler2015generic}
\begin{equation}
	B(\tau)=\sum_{i,j=1, i\neq j}^{N}\left<b_{ij}(t_0) b_{ij}(\tau)   \right>,
\end{equation}

where the brackets $\left<..\right>$ denote an average over all the particle pairs that are bonded at time $t_0$. 

\section{Position-resolved local area fractions} \label{appendix: position resolved local area fracs}
To study the occurrence of phase separation, we calculate the position-resolved local area fraction distributions. A phase-separated state is indicated by a double peak structure of this distribution. We first begin by calculating the local area fraction $\phi_i$ around each particle $i$. To this end, a Voronoi cell is constructed around each particle $i$ whose area is denoted by $A_i$ \cite{okabe2009spatial}. In order to ensure that the central simulation box is properly partitioned by the Voronoi tessellation algorithm, we also take into account four nearest images of the centrally placed actual simulation box \cite{liao2018clustering}. The local area fraction of each particle is then given by

\begin{equation}
	 \phi_i=\frac{\text{area occupied by particle i}}{\text{area of the corresponding Voronoi cell}}=\frac{\pi(\sigma/2)^2}{A_i}.
\end{equation}

Now, we divide the system into a grid where the mesh size is set to be equal to $\sigma$, such that it is large enough to preserve the particle-resolved information \cite{liao2018clustering, blaschke2016phase}. For each grid point $(x,y)$ we first identify the index $i$ of the Voronoi cell that it falls within and assign $\phi(x,y)=\phi_i$, to finally obtain the position-resolved local area fractions. We further perform a time average ($\bar{\phi}(x,y)$) over the last 30 simulation time points to filter out any transient small clusters in the dilute region \cite{blaschke2016phase, liao2018clustering}. From this, we finally obtain a histogram representing the probability distribution of $\bar{\phi}(x,y)$. 

For obtaining the coexistence binodal, it is important to correctly identify peaks that correspond to the dense clusters and the dilute regions in the system from the probability distribution of $\bar{\phi}(x,y)$. This can prove to be rather challenging at certain parameter combinations of $\delta_r$ and $\kappa_{\text{nr}}$. It is important to note that for self-aggregating systems that form a single large cluster, the particles at the edges of the clusters can show a low value of $\phi_i$. Since we perform numerical simulations at a reduced number density of $\rho^*=0.3$, this can lead to a lot of empty space around the cluster(s). In such cases, the particles at the edges of the cluster show a false peak in the probability distribution at around $\bar{\phi}(x,y)\leq0.01$. To avoid this effect, we detect the particles at the edges of clusters and discard those Voronoi cells which are mapped to them. Such a process ensures that peaks observed at low values of $\bar{\phi}(x,y)$ correspond to an actual dilute phase and are not caused as a result of particles at the edges of clusters.

\section{Cluster centre-of-mass velocities}  \label{appendix: clus_vels}
Assume that we have a cluster of $N_{cl}$ particles. The position vector of the centre-of-mass at time $t$ [$\mathbf{R}_{cm}(t)$] is given by 
\begin{equation}
	\mathbf{R}_{cm}(t)=\frac{1}{N_{cl}}\sum_{i=1}^{N_{cl}}\mathbf{r}_i(t). 
\end{equation} 

Here, $\mathbf{r}_i(t)$ are the position vectors of each particle $i$ belonging to the cluster at time $t$. Now, the translational velocity of the centre-of-mass at time $t$ can be calculated using $\mathbf{R}_{cm}(t)$ and $\mathbf{R}_{cm}(t+\Delta t)$, i.e., 
\begin{equation}
	\mathbf{v}(t)=\frac{\mathbf{R}_{cm}(t+\Delta t)-\mathbf{R}_{cm}(t)}{\Delta t},
\end{equation} 

where $\Delta t$ is the time interval considered. In order to calculate the angular velocity, we first calculate the relative position vectors of each of the constituent particles from the centre of mass at time $t$,
\begin{equation}
	\mathbf{r}_i^{rel}(t)=\mathbf{r}_i(t)-\mathbf{R}_{cm}(t). 
\end{equation}

The angle made by $\mathbf{r}_i^{rel}(t)$ and the $x-$ axis is given by $\theta_i(t)=\cos^{-1}\left(\mathbf{\hat{x}}.\mathbf{r}_i^{rel}(t)/|\mathbf{\hat{x}}| |\mathbf{r}_i^{rel}(t)|\right)$. Finally, the angular velocity is then given by 
\begin{equation}
	\omega(t)=\frac{1}{N_{cl}}\sum_{i=1}^{N_{cl}}\frac{\theta_i(t+\Delta t)- \theta_i(t)}{\Delta t}. 
\end{equation} 

The quantities $\mathbf{v}(t)$ and $\omega(t)$ are calculated from the simulation trajectory data every $10^{-3}\tau_B$ within the time period that the cluster remains stable. Finally, the time-averaged translational velocity magnitude $|\mathbf{\bar{v}}|$ and rotational velocity $\bar{\omega}$ can be calculated. The results are given in Table \ref{Table:clus_vels}.

\section{Calculation of the characteristic domain size} \label{appendix: length_scale}
The characteristic domain size $L(t)$ is calculated by identifying the position of the peak in the structure factor $S(\mathbf{k})$ defined as,
\begin{equation}
	S(\mathbf{k})=\frac{1}{N}\left< \rho(\mathbf{k})\rho(-\mathbf{k}) \right>.
\end{equation}

Here, $\rho(\mathbf{k})=\int \rho(\mathbf{r})e^{-i\mathbf{k.r}}d\mathbf{r}$ is the Fourier transform of the density field $\rho(\mathbf{r})=\sum_{j=1}^{N}\delta(\mathbf{r}-\mathbf{r}_j)$. We then identify the wave-vector ($\mathbf{k}_{\text{max}}$) corresponding to the peak in $S(\mathbf{k})$  and finally obtain the domain size $L(t)=\frac{2\pi}{|\mathbf{k}_\text{max}|}$.

\bibliography{aiptemplate}% Produces the bibliography via BibTeX.

\end{document}